\documentclass[useAMS,usenatbib]{mn2e}
\usepackage{graphicx}
\usepackage{times}
\usepackage{amssymb}
\usepackage{amsmath}
\usepackage{multirow}
\usepackage{subfig}
\usepackage{hyphenat}
\newcommand{\hompc}{\,h\,{\rm Mpc}^{-1}}
\newcommand{\mpcoh}{\,h^{-1}\,{\rm Mpc}}

\begin{document}

\title[Mock catalogues \& growth rate measurement at $z=0.15$] 
{The Clustering of the SDSS Main Galaxy Sample II: Mock galaxy catalogues and a measurement of the growth of structure from Redshift Space Distortions at $z=0.15$}

\author[C. Howlett et al.]{\parbox{\textwidth}{
Cullan Howlett\thanks{Email: cullan.howlett@port.ac.uk}$^{1}$, 
Ashley J. Ross$^{1}$,
Lado Samushia$^{1,2,3}$,
Will J. Percival$^{1}$,
Marc Manera$^{4,1}$
}
  \vspace*{4pt} \\ 
$^{1}$Institute of Cosmology \& Gravitation, Dennis Sciama Building, University of Portsmouth, Portsmouth, PO1 3FX, UK\\
$^{2}$Department of Physics, Kansas State University, 116, Cardwell Hall, Manhattan, KS, 66506, USA \\
$^{3}$National Abastumani Astrophysical Observatory, Ilia State University, 2A Kazbegi Ave., GE-1060 Tbilisi, Georgia\\
$^{4}$University College London, Gower Street, London WC1E 6BT, UK\\
}
\date{draft} 

\pagerange{\pageref{firstpage}--\pageref{lastpage}} \pubyear{2014}
\maketitle
\label{firstpage}

\begin{abstract}

We measure Redshift-Space Distortions (RSD) in the two-point correlation function of a sample of $63,163$ spectroscopically identified galaxies with $z < 0.2$, an epoch where there are currently only limited measurements, from the Sloan Digital Sky Survey (SDSS) Data Release 7 Main Galaxy Sample. Our sample, which we denote MGS, covers 6,813 deg$^2$ with an effective redshift $z_{eff}=0.15$ and is described in our companion paper (Paper I), which concentrates on BAO measurements. In order to validate the fitting methods used in both papers, and derive errors, we create and analyse 1000 mock catalogues using a new algorithm called {\sc picola} to generate accurate dark matter fields. Haloes are then selected using a friends-of-friends algorithm, and populated with galaxies using a Halo-Occupation Distribution fitted to the data. 
Using errors derived from these mocks, we fit a model to the monopole and quadrupole moments of the MGS correlation function. If we assume no Alcock-Paczynski (AP) effect (valid at $z=0.15$ for any smooth model of the expansion history), we measure the amplitude of the velocity field, $f\sigma_{8}$, at $z=0.15$ to be $0.49_{-0.14}^{+0.15}$. We also measure $f\sigma_{8}$ including the AP effect. This latter measurement can be freely combined with recent Cosmic Microwave Background results to constrain the growth index of fluctuations, $\gamma$.  Assuming a background $\Lambda$CDM cosmology and combining with current Baryon Acoustic Oscillation data we find $\gamma = 0.64 \pm 0.09$, which is consistent with the prediction of General Relativity ($\gamma \approx 0.55$), though with a slight preference for higher $\gamma$ and hence models with weaker gravitational interactions.

\end{abstract}


\section{Introduction}

The observed 3D clustering of galaxies provides a wealth of cosmological information: the comoving clustering pattern was encoded in the early Universe and thus depends on the physical energy densities (e.g. \citealt{Pee70,Sun70,Dor78}), while the bias on large-scales encodes primordial non-Gaussianity \citep{dalal08}. Secondary measurements can be made from the observed projection of this clustering, including using Baryon Acoustic Oscillations (BAO) as a standard ruler \citep{seo03,blake03} or by comparing clustering along and across the line-of-sight \citep{AP}. In this paper we focus on a third type of measurement that can be made, called Redshift-Space Distortions (RSD; \citealt{Kai87}). RSD arise because redshifts include both the Hubble expansion, and the peculiar velocity of any galaxy. The component of the peculiar velocity due to structure growth is coherent with the structure itself, leading to an enhanced clustering signal along the line-of-sight. The enhancement to the overdensity is additive, with the extra component dependent on the amplitude of the velocity field, which is commonly parameterised on large-scales by $f\sigma_{8}$, where $f\equiv d\ln D/d\ln a$ is the logarithmic derivative of the growth factor with respect to the scale factor and $\sigma_{8}$ is the linear matter variance in a spherical shell of radius $8\mpcoh$. Together these parameterise the amplitude of the velocity power spectrum. 

The largest spectroscopic galaxy survey undertaken to-date is the Sloan Digital Sky Survey (SDSS), which has observed multiple samples over its lifetime. The SDSS-I and SDSS-II \citep{York00} observed two samples of galaxies: the $r$-band selected main galaxy sample \citep{Strauss02}, and a sample of Luminous Red Galaxies (LRGs; \citealt{Eis01}) to higher redshifts. The Baryon Oscillation Spectroscopic Survey (BOSS; \citealt{Daw12}), part of SDSS-III \citep{Eis11} extended the LRG sample to higher redshifts with a sample at $z\sim0.57$ called CMASS, and a sample at $z\sim0.32$ called LOWZ that subsumed the SDSS-II LRG sample. SDSS-IV will extend the LRG sample to even higher redshifts, while simultaneously observing a sample of quasars and Emission Line Galaxies (ELGs).

In this paper we revisit the SDSS-II main-galaxy sample, herein denoted MGS, applying the latest analysis techniques. We have sub-sampled this catalogue to select high-bias galaxies at $z<0.2$ (details can be found in our companion paper \citealt{Ross14}, Paper I, which also presents BAO-scale measurements made from these data). This sampling positions the galaxies redshift between BOSS LOWZ, and the 6-degree Field Galaxy Survey (6dFGS; \citealt{Beutler11}), filling in a gap in the chain of measurements at different redshifts. Selecting high-bias galaxies means that we can easily simulate the sample. In this paper we present RSD measurements made using the MGS data. 

Recent analyses of BOSS have emphasised the importance of accurate mock catalogues \citep{Manera2013,Manera2014}; these provide both a mechanism to test analysis pipelines and to determine covariances for the measurements made. For the MGS data, we create 1000 new mock catalogues using a fast N-body code based on a new parallelisation of the {\sc cola} algorithm \citep{Tassev2013}, designed to quickly create approximate evolved dark matter fields. Haloes are then selected using a friends-of-friends algorithm, and a Halo-Occupation Distribution based method is used to populate the haloes with galaxies. The algorithms and methods behind {\sc picola} can be found in Howlett et. al. (in prep.).

Our paper is outlined as follows: In Section 2 we describe the properties of the MGS data. In Section 3 we summarise how we create dark matter halo simulations using {\sc picola}. In Section 4, we describe how we calculate clustering statistics, determine the halo occupation distribution we apply to mock galaxies to match the observed clustering, and test for systematic effects. In Section 5, we describe how we model the redshift space correlation function using the Gaussian Streaming/Convolved Lagrangian Perturbation Theory (CLPT) model of \citet{Wang2014}. In Section 6, we describe how we fit the MGS clustering in the range $25 \mpcoh \le s \le 160 \mpcoh$, test our method and validate our choice of fitting parameters and priors using the mock catalogues. In Section 7 we present the results from fitting to the MGS data and present our constraints on $f\sigma_{8}$. In Section 8, we compare our measurements to RSD measurements at other redshifts, including results from \cite{Beutler12,Chuang2013,Samushia2012} and \cite{Samushia2014}, and test for consistency with General Relativity. We conclude in Section 9. Where appropriate, we assume a fiducial cosmology given by $\Omega_m = 0.31$, $\Omega_b=0.048$, $h=0.67$, $\sigma_{8}=0.83$, and $n_s=0.96$.

\section{Data}
\label{sec:data}
\subsection{The Completed SDSS Main Galaxy Sample}

We use the same SDSS DR7 MGS data as analysed in Paper I, which is drawn from the completed data set of SDSS-I and SDSS-II. These surveys obtained wide-field CCD photometry (\citealt{C,Gunn06}) in five passbands ($u,g,r,i,z$; \citealt{F}), amassing a total footprint of 11,663 deg$^2$, internally calibrated using the `uber-calibration' process described in \cite{Pad08}, and with a 50\% completeness limit of point sources at $r = 22.5$ (\citealt{DR7}). From these imaging data, the main galaxy sample (MGS; \citealt{Strauss02}) was selected for spectroscopic follow-up, which to good approximation, consists of all galaxies with $r_{\rm pet} < 17.77$, where $r_{\rm pet}$ is the extinction-corrected $r$-band Petrosian magnitude, within a footprint of 9,380 deg$^2$ \citep{DR7}.

\begin{figure}
\includegraphics[width=84mm]{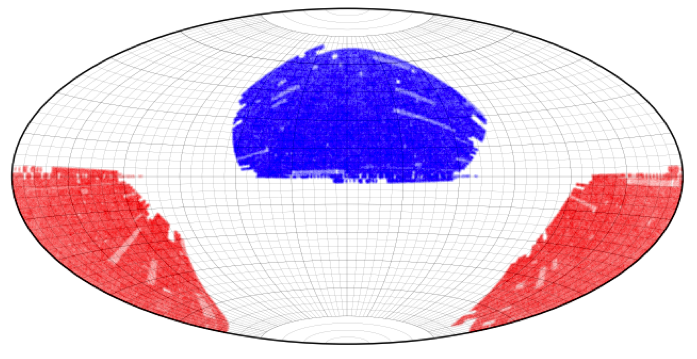}
  \caption{The blue area shows a flat, all-sky projection of the footprint of our MGS sample, which occupies 6,813 deg$^2$. The red area shows the same geometry, after a 180$^{\rm o}$ rotation. This illustrates how we produce two mock galaxy samples from every full-sky dark matter halo catalog.}
  \label{fig:foot}
\end{figure}

For our analysis, we start with the SDSS MGS value-added galaxy catalog `safe0'  hosted by NYU\footnote{http://sdss.physics.nyu.edu/vagc/lss.html} (NYU-VAGC), which was created following the methods described in \cite{Blanton05}. The catalog includes $K$-corrected absolute magnitudes, determined using the methods of \cite{Blanton03}, and detailed information on the mask. We only use the contiguous area in the North Galactic cap and only areas where the completeness is greater than 0.9, yielding a footprint of 6,813 deg$^2$, compared to the original 7,356 deg$^2$. We create the mask describing this footprint from the window given by the NYU-VAGC, which provides the completeness in every mask region, and the {\sc Mangle} software \citep{Swanson2008}.  We also use the {\sc Mangle} software to obtain angular positions of unclustered random points, distributed matching the completeness in every mask region. The angular footprint of our sample is displayed in blue in Fig.~\ref{fig:foot}. The red patch in Fig.~\ref{fig:foot} shows the angular footprint of our galaxy sample after rotating the coordinates via $RA \Rightarrow RA+\pi$,  $DEC   \Rightarrow -DEC$ and once again applying the mask. As described in Section \ref{sec:sim}, we choose to create full-sky simulations, and in doing so, we can use the mask to create two mock galaxy catalogues that match our footprint, reducing the noise in our estimates of the covariance matrix at almost no extra cost\footnote{In principle, we could fit $\sim 6$ replicates of our survey footprint in each full-sky simulation without overlap, though not, perhaps, without significant cross-correlation between patches taken from the same realisation. In practice we simply generate two survey patches from each simulation.}.

\begin{figure}
\includegraphics[width=84mm]{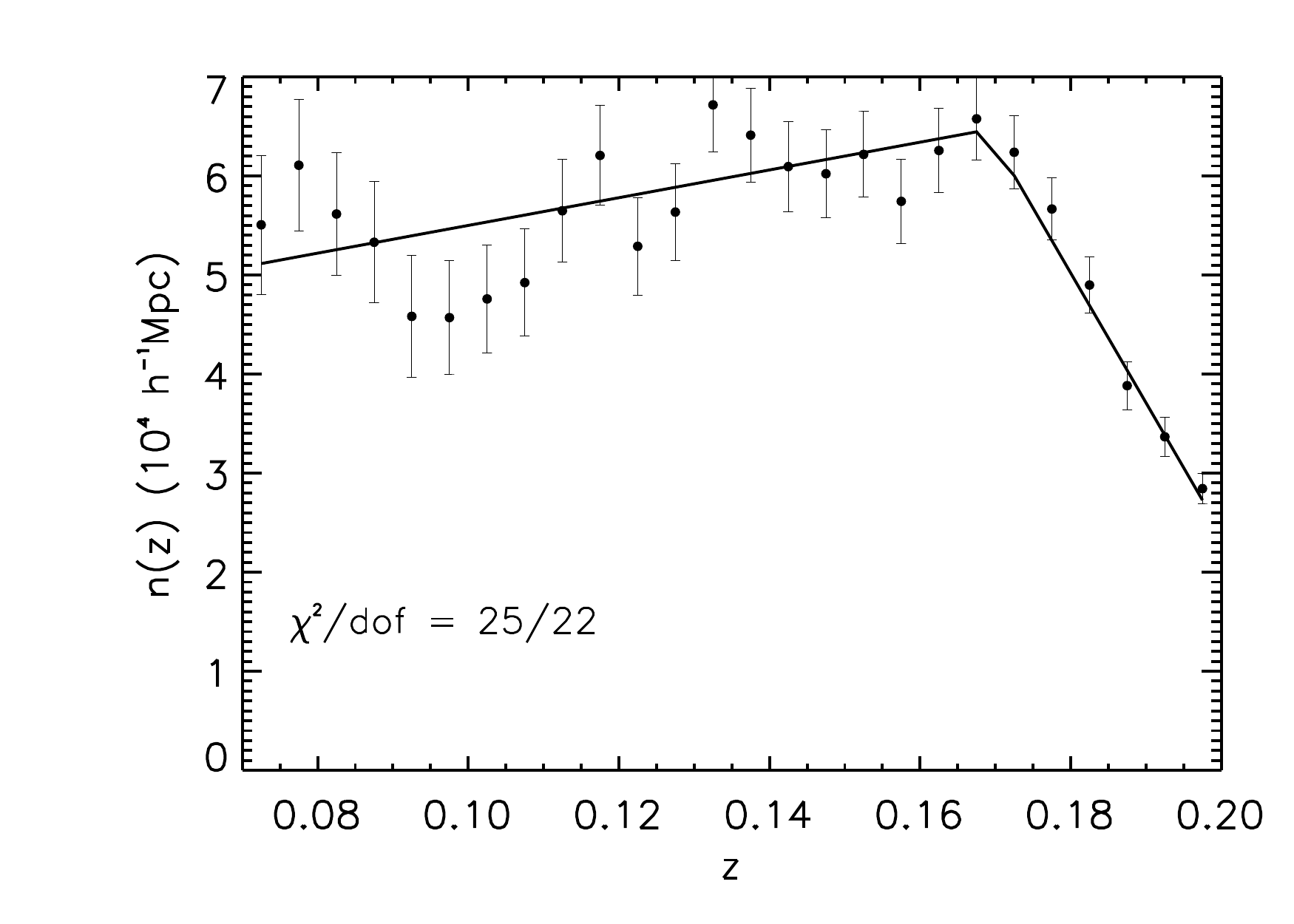}
  \caption{The number density as a function of redshift for our galaxy sample compared to the mean of the mocks after subsampling. The error bars come from the standard deviation of our 1000 mock realisations}
  \label{nzmock}
\end{figure}

We make further cuts on the NYU VAGC safe0 sample based on colour, magnitude, and redshift. These are $0.07 < z < 0.2$, $M_r < 21.2$ and $g-r > 0.8$, where $M_r$ is the $r$-band absolute magnitude provided by the NYU-VAGC. These cuts produce a sample of moderately high bias ($b\sim1.5$), with a nearly constant number density that is independent of BOSS and 6dFGS samples. The resulting sample contains 63,163 galaxies. The redshift distribution is shown in Fig.~\ref{nzmock}. The effective redshift of our sample is $z_{\rm eff} = 0.15$, calculated as described in Paper I, where further details on the sample selection criteria can be found. 

Fig.~\ref{nzmock} also shows (solid line) the average number density of the mock galaxy catalogues described in Section~\ref{sec:sim}. We determine the $n(z)$ that we apply to the mocks by fitting to a model with two linear relationships and a transition redshift. The best-fit is given by
\begin{equation}
n(z) = 
  \begin{cases}
    0.0014z+0.00041; & z < 0.17 \\
    0.00286-0.0131z; & z\geq 0.17 \label{eq:nz}.
  \end{cases}
 \end{equation}
We see that the mock galaxy catalogues agree with the data very well, with $\chi^{2} =25$ for $22$ degrees of freedom (26 redshift bins and 4 independent fitting parameters). The errors come from the standard deviation in number density across the set of mock catalogues.

\section{Simulations}
\label{sec:sim}
Simulations of our MGS data are vital in order to accurately estimate the covariance matrix of our clustering measurements and to perform systematic tests on our BAO and RSD fitting procedures. Of order 1000 mock galaxy catalogues (mocks) are necessary to ensure noise in the covariance matrix does not add significant noise to our measurements \citep{Per14}. For BOSS galaxies, such mocks were created using the methods described in \cite{Manera2013, Manera2014}. The galaxies in our sample have lower bias than those of BOSS, and we therefore require a method of producing dark matter halos at higher resolution than used in BOSS, yet in such a way that we can still create a large number of realisations in a timely fashion. For this we have created the code {\sc picola}, a highly-developed, planar-parallel implementation of the COLA method of \cite{Tassev2013}; this implementation is described in Howlett et. al. (in prep.), and a user guide that will be included with the public release of the code. It should be noted that a similar method has also recently been used to create mock catalogues for the WiggleZ survey \citep{Kazin2014}, though the codes were developed independently.

In this section, we describe how we use {\sc picola} to produce dark matter fields and then halo catalogues, and how we apply a Halo Occupation Distribution (HOD, \citealt{Berlind2002}) prescription to these halo catalogues to produce mock galaxy catalogues. We expect that the methods we use to generate these halo catalogues will be generally applicable to any future galaxy survey analyses. In Section \ref{sec:clus}, we describe how we specifically fit an HOD model to the measured clustering of the MGS to produce mocks that simulate our MGS data. These mocks are used in the RSD analyses we present and the BAO analysis of Paper~I.  

\subsection{Producing Dark Matter fields}

We generate 500 dark matter snapshot realisations using our fiducial cosmology, which we convert into 1000 mock galaxy catalogues. Although our code is capable of generating lightcones `on the fly' without sacrificing speed, we stick with snapshots for simplicity in later stages and because we expect the inaccuracies arising from using snapshots to be small due to the low redshift of our sample. For each simulation we evolve $1536^{3}$ particles, with a mesh size equal to the mean particle separation, in a box of edge length $1280\mpcoh$. We choose this volume as it is large enough to cover the full sky out to the maximum comoving distance of our sample at $z=0.2$ (for our fiducial cosmology this is $\sim 570\mpcoh$). We evolve our simulation from $z=9.0$ to $z=0.15$, using 10 timesteps equally spaced in $a$, the scale factor. This results in a mass resolution of $\sim 5\times10^{10}\,h^{-1}\,\rm{M}_{\odot}$, a factor of $10$ smaller than that used for the BOSS LOWZ mock catalogues. Each simulation takes around 20 minutes (including halo-finding) on 256 cores. In terms of the actual computing time used, our {\sc picola} run took $\sim 25$ CPU-hours compared to $\sim 27600$ CPU-hours for the {\sc gadget-2} run described below. However, it should be noted that the actual (wall)time taken for the {\sc gadget-2} run was not 1000 times that of the {\sc picola} run, rather the memory requirements of {\sc gadget-2}  are also larger than those of {\sc picola}, requiring more processors to run (384 in this case).

\begin{figure}
\includegraphics[width=84mm]{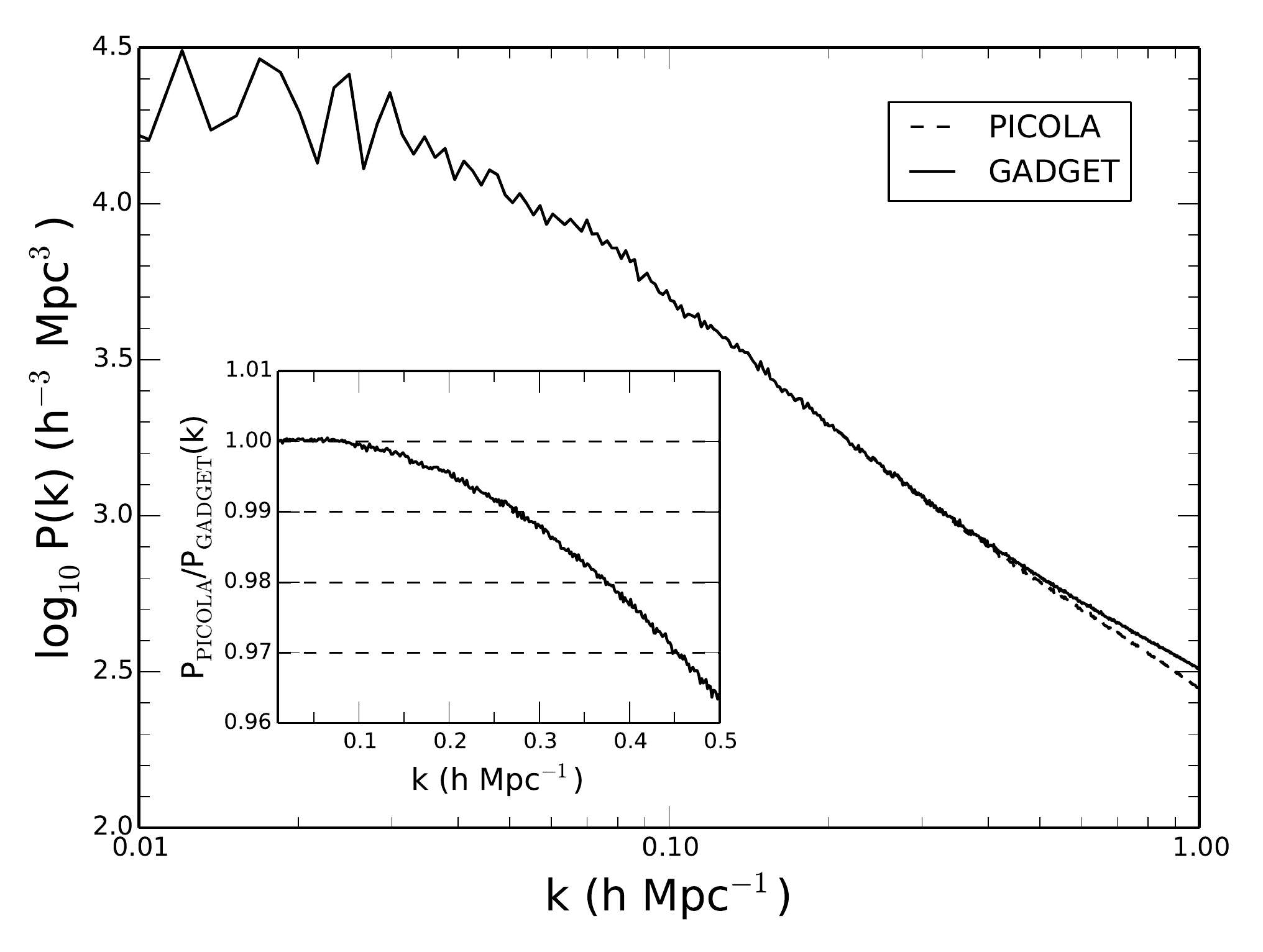}
  \caption{The power spectrum of the dark matter field in a cubic box from the {\sc picola} and {\sc gadget-2} runs described in the text. We can see good agreement between the two even into the non-linear regime.}
  \label{pkpicola}
\end{figure}

Fig.~\ref{pkpicola} shows the power spectrum of the dark matter fields for one of our {\sc picola} simulations and for a Tree-PM N-Body simulation performed using {\sc gadget-2} \citep{Springel2005}. Both simulations use the same initial conditions and the same mesh resolution. We can see that the power spectra agree to within 2 percent across all scales of interest to BAO measurements and the agreement continues to within 10 percent to $k \sim 0.8\hompc$.

\subsection{From Dark matter to Halos}

We generate halos for our {\sc picola} dark matter simulations using the friends-of-friends algorithm (FoF; \citealt{Davis1985}) with linking length equal to the commonly used value of $b=0.2$, in units of the mean particle separation. We average over all of the constituent particles of each halo to calculate the position and velocity of the centre-of-mass. The halo mass, $M$, is given by the individual particle mass multiplied by the number of constituent particles that make up the halo. The virial radius is then estimated as
\begin{equation}
R_{vir} = \left( \frac{3M}{4\pi \rho_{c}(z) \Delta_{vir} \Omega_{m}(z)} \right)^{1/3},
\end{equation}
where $\rho_{c} \approx 2.77\times10^{11}\,h^{2}\,{\rm M_{\odot}}\, {\rm Mpc}^{-3}$ is the critical density, and we use a value $\Delta_{vir}=200$ (e.g. \citealt{Tinker2008}).

\begin{figure}
\includegraphics[width=84mm]{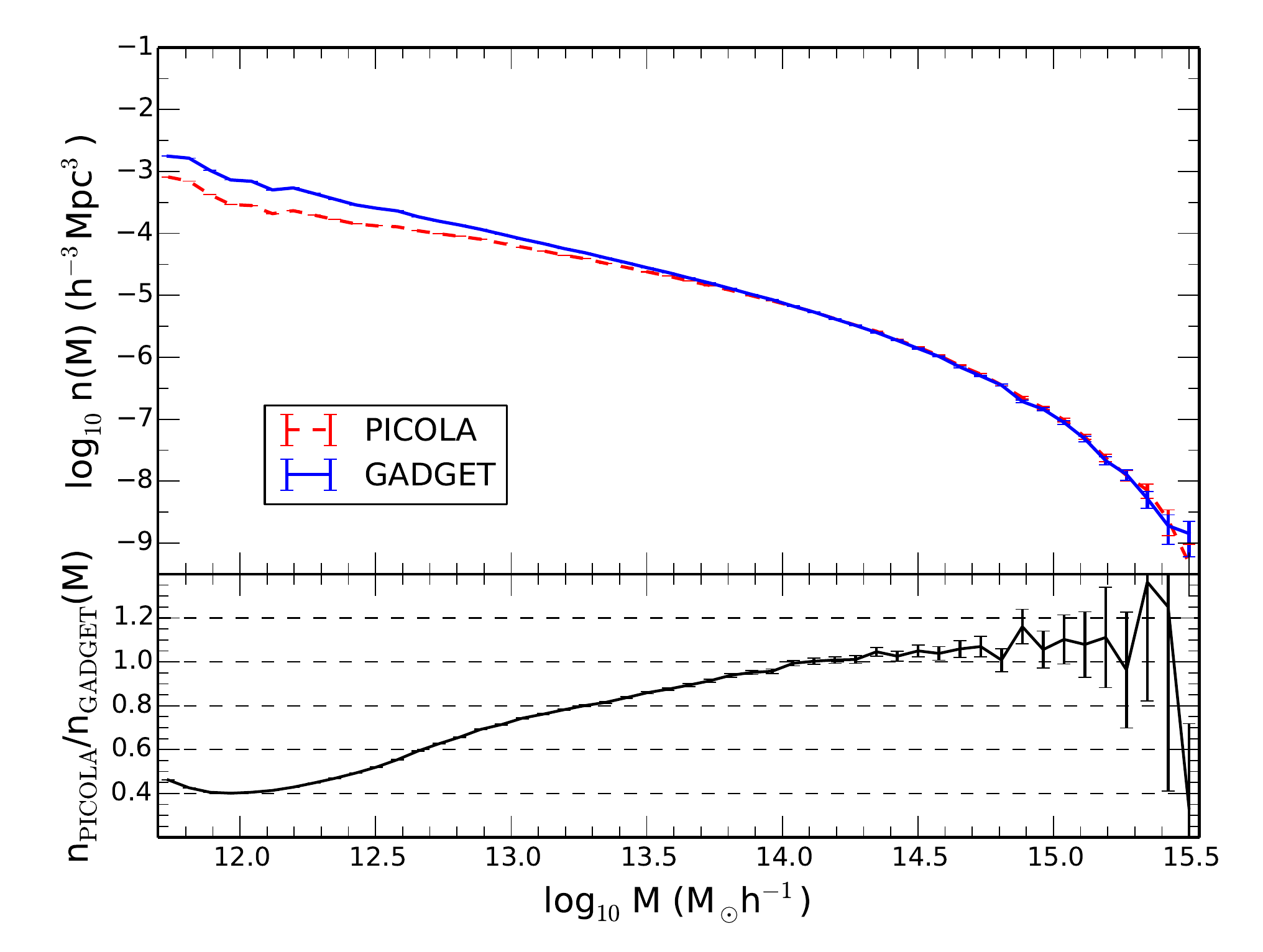}
  \caption{A comparison of the halo mass function from our {\sc gadget-2} and {\sc picola} simulations run from the same initial conditions. We see a lack of halos on small scales due to the finite mesh resolution, but this is easily compensated for with the HOD fitting described later.}
  \label{halomass}
\end{figure}

The clustering of the dark matter particles is recovered well by {\sc picola}. It is slightly under-represented on small scales, but we do not need to modify the linking length in order to recover our halos (unlike, for example, in \citealt{Manera2013}). Fig.~\ref{halomass} shows the level of agreement between halo mass functions recovered from our matched parameter {\sc picola} and {\sc gadget-2} runs. The difference in halo number density for low-mass halos is a direct consequence of the mesh resolution of our simulations. As {\sc picola} does not calculate additional contributions to the inter-particle forces (i.e., via a Tree-level Particle-Particle summation) on scales smaller than the mesh, using instead the approximate, interpolated forces from the nearest mesh points, we do not produce the correct structure on the order of a few mesh cells or smaller. This results in slightly `puffy' halos.

\begin{figure}
\includegraphics[width=84mm]{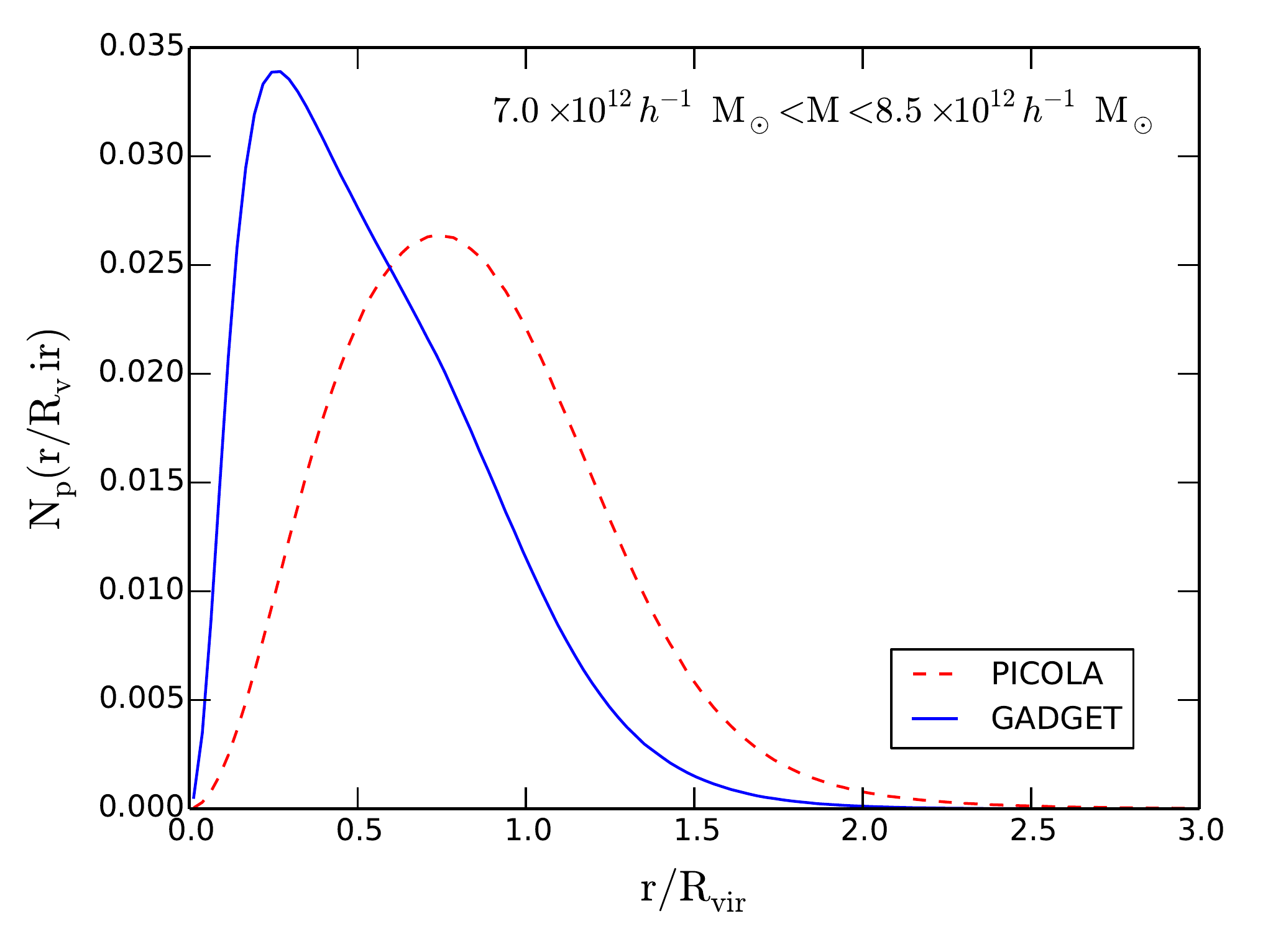}
  \caption{The normalised number of constituent dark matter particles found within a halo as a function of their separation from the halo centre of mass, in units of the virial radius, for a given halo mass range. We see that the halos from {\sc picola} are generally more dispersed than those from {\sc gadget-2}, where the particles have not collapsed sufficiently for the FoF algorithm to group them. This in turn leads to a slight lack of low mass halos overall, which we are able to correct for in our HOD fitting method.}
  \label{PICOLAdensity}
\end{figure}

This is shown in Figure~\ref{PICOLAdensity}, where for halos within a given mass range we plot the normalised number of dark matter particles in that halo as a function of their separation from the centre of mass, normalised by the halo virial radius. For the halo mass range in question we see that the constituent particles of the PICOLA halos are located at slightly larger radii than their GADGET counterparts. This difference is reduced as we go to higher mass halos where the overall properties of the halo are still captured. However, it does mean that we miss some of the outlying particles of the larger halos, and some smaller halos altogether, as the dark matter particles have not collapsed sufficiently to be grouped together by the FoF algorithm.

Regardless of this, the effect is small enough over the halo mass range of interest for the MGS that we find no correction is necessary before we apply our HOD model. In addition, as described in Section~\ref{sec:PHOD}, we determine the HOD parameters directly by populating mock dark matter halos. The deficit of lower mass halos is thus compensated for by assigning more galaxies to lower mass halos. It should also be noted that although other halo-finding techniques may produce better results, we retain the FoF algorithm in the interest of speed.

\subsection{Assigning Galaxies to Halos}

We populate our halos in a very similar way to that of \cite{Manera2013} using the HOD model \citep{Berlind2002}. Within this framework we assign galaxies to halos based solely on the mass of the halo, splitting the galaxies into central and satellite types. We define two mass-dependent functions, $\langle N_{cen}(M) \rangle$ and $\langle N_{sat}(M) \rangle$, where $\langle N_{cen}(M) \rangle$ denotes the probability that a halo of mass $M$ contains a central galaxy and $\langle N_{sat}(M) \rangle$ is the mean of the poisson distribution from which we randomly generate the number of satellite galaxies. These functions are themselves modelled with parameters estimated from a fit to the MGS data, as described in Section~\ref{sec:PHOD}.

Central galaxies are placed at the centre of mass of the halo, and satellites at radii $r \le R_{vir}$ with probability derived from the NFW profile \citep{NFW}
\begin{equation}
\rho(r) = \frac{4\rho_{s}}{\frac{r}{r_{s}}\left(1+\frac{r}{r_{s}}\right)^{2}},
\label{nfw}
\end{equation}
where $r_{s}=R_{vir}/c_{vir}$ is the characteristic radius, at which the slope of the density profile is -2, and $\rho_{s}$ is the density at this radius. $c$ is the concentration parameter, which we calculate for a halo of mass $M$ using the fitting formulae of \cite{Prada2012}. On top of this we add a dispersion to the mass-concentration relation using a lognormal distribution with mean equal to that evaluated from the fitting functions and variance $\sigma=0.078$. This is the same value as that used in \cite{Manera2013} and is a typical value, as measured from fitting NFW profiles to halos recovered from simulations \citep{Giocoli2010}.

Both central and satellite galaxies are given the velocity of the centre of mass of the halo. Satellite galaxies are then assigned an extra peculiar velocity contribution drawn from a Gaussian, with the velocity dispersion calculated from the virial theorem
\begin{equation}
\langle v^{2} \rangle = \left\langle \frac{GM(r)}{r} \right\rangle.
\end{equation}
For an NFW profile, the mass inside a radius $r$ is
\begin{equation}
M(r)=4\pi\rho_{s}r_{s}^{3}\left[\text{ln}\left(\frac{r_{s}+r}{r_{s}}\right)-\frac{r}{r_{s}+r}\right],
\end{equation}
and hence the velocity dispersion for a halo of mass $M$ is
\begin{equation}
\langle v^{2} \rangle = \frac{GM}{r_{s}} \frac{c(1+c)-(1+c)\text{ln}(1+c)}{2((1+c)\text{ln}(1+c)-c)^{2}}.
\end{equation}
In order to assign the additional satellite velocities in each direction we use a gaussian distribution with zero mean and variance $\langle v^{2} \rangle/3$.

To simulate the effects of Redshift-Space Distortions we displace each galaxy along the line-of-sight by
\begin{equation}
\Delta s_{los} = \frac{v_{los}}{H(z)a},
\end{equation}
Given $\Delta s_{los}$ and a galaxy's true position, we determine angles and redshifts using our fiducial cosmology, placing the observer at the centre of each simulation box. 

\section{Clustering}
\label{sec:clus}

\subsection{Power Spectrum}
\label{sec:Pk}

\begin{figure}
\includegraphics[width=84mm]{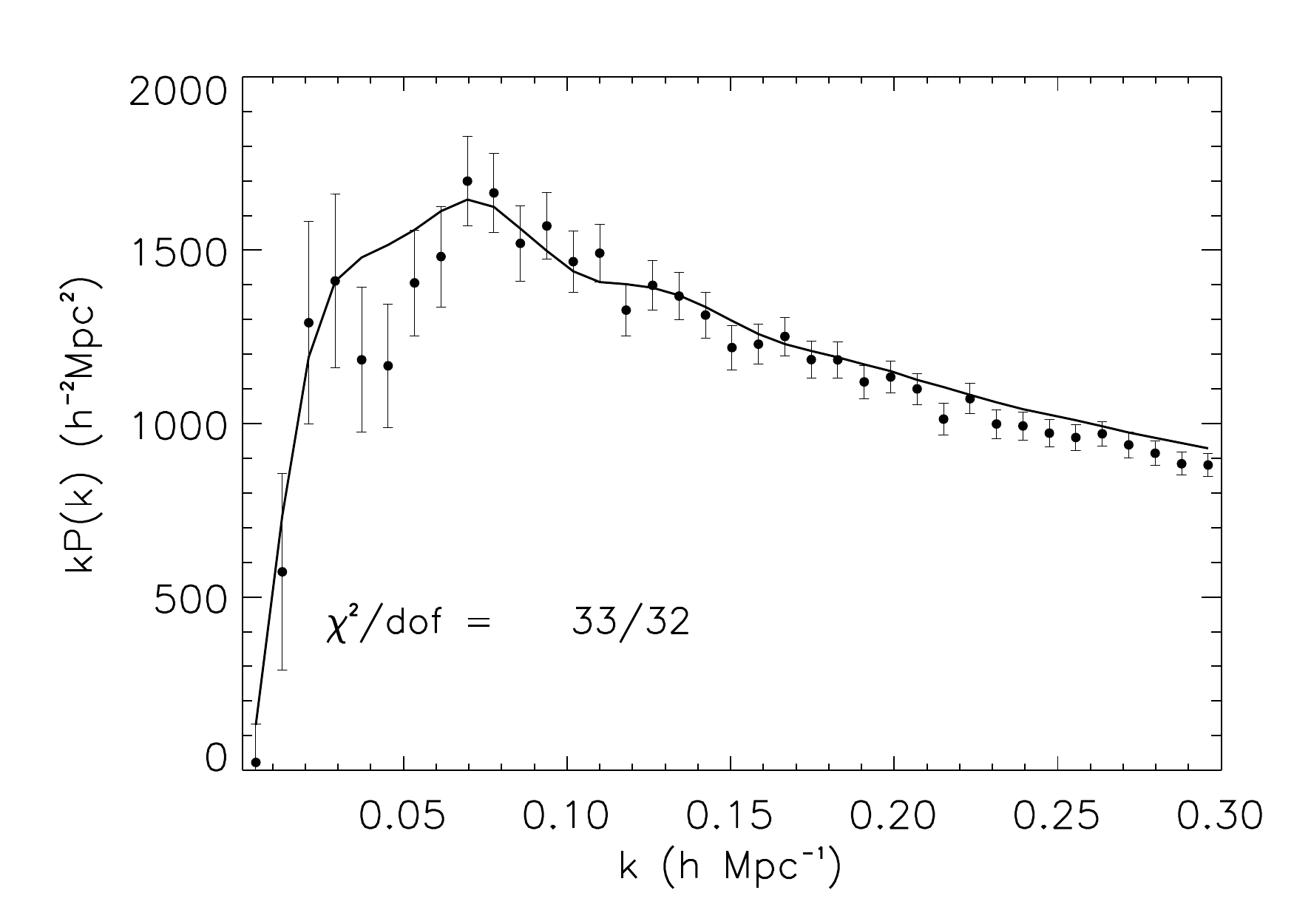}
  \caption{The power spectrum of our sample. Points show the data and the solid line shows the mean of the mocks. The error bars come from the diagonal elements of the covariance matrix constructed using the mock catalogues.}
  \label{pkmock}
\end{figure}

Although we obtain our cosmological constraints from measuring the correlation function and not the data power spectrum, we do use the monopole moment of the power spectrum to determine the HOD model used for the mocks, as it is faster to compute than its configuration-space analogue. We estimate the monopole of the power spectrum, which we denote $P(k)$, using the Fourier-based method of \cite{FKP}. We convert each galaxy's redshift space coordinates to a cartesian basis using our fiducial cosmology. We then compute the overdensity on a grid containing $1024^{3}$ cells in a box of edge length $2000\mpcoh$. This provides ample room to zero pad the galaxies to improve the frequency sampling and results in a Nyquist frequency of $1.6\hompc$, much larger than the largest frequency of interest. We use the random catalogue to estimate the expected density at each grid point. Galaxies and randoms are weighted based on the number density as a function of redshift,
\begin{equation}
w_{FKP}(z) = \frac{1}{1+n(z)P_{FKP}}
\label{weights}
\end{equation}
where we set $P_{FKP} = 16000\,h^{-3}\rm{Mpc}^{3}$, which is close to the measured amplitude at $k=0.1\hompc$. This corresponds to physical scales $\sim 60 \mpcoh$, which are well within our fitting range, and, in any case, the efficiency of this weighting system has only a very weak scale-dependence. After Fourier transforming the overdensity grid we calculate the spherically-averaged power spectrum in bins of $\Delta {k} = 0.008$, correcting for gridding effects and shot-noise. The power spectrum of the MGS data is displayed as points in Fig.~\ref{pkmock}. The smooth curve and error-bars display the mean of the mock $P(k)$ and their standard deviation.

\subsection{HOD fitting}
\label{sec:PHOD}

We match the measured $P(k)$ of the MGS and the average from 10 halo catalogues in order to determine the HOD model that we then apply to all of the mock catalogues. In this way, we do not need to correct our halo mass function at the low-mass end, as the lack of low-mass halos will be compensated via the population of lower-mass halos. 

We use the five parameter functional form of \cite{Zheng2007} for the number of central and satellite galaxies, 
\begin{align}
  \langle N_{cen}(M)\rangle &= \dfrac{1}{2}\left [1 + \text{erf}\left (\dfrac{\text{log}M -\text{ log}M_{min}}{\sigma_{\text{log}M}}\right )\right ], \notag\\[9pt]
  \langle N_{sat}(M)\rangle &= \langle N_{cen}\rangle \left (\dfrac{M-M_{cut}}{M_{1}}\right )^\alpha.
\end{align}
For a halo of $M < M_{cut}$ we set $\langle N_{sat}\rangle=0$ and in the case where we assign satellite galaxies but no central galaxy to a halo, we remove one of the potential satellite galaxies and replace it with a central. We set the values of the five free parameters by iterating over the following steps:
\begin{enumerate}
  \item{Populate a subset of the mocks using a given set of HOD parameters,}
  \item{Mask the mock galaxies so that they match the data,}
  \item{Subsample the mock galaxies to match our idealised $n(z)$,}
  \item{Calculate the average power spectrum of our populated mocks and compare to the data.}
\end{enumerate}
We use 10 mocks to fit our HOD, populating and masking them individually, but reproducing the radial selection function by sub-sampling based on the ratio between the analytic fit to the data $n(z)$ and the average number density of the 10 mocks. The fit is performed using a downhill simplex minimisation of the $\chi^{2}$ difference between the average, 10-mock power spectrum and the data power spectrum in the range $0.02 \le k \le 0.3$. The fit is performed twice, first using analytic errors on the power spectrum from \cite{Tegmark1997} (equations 4 and 5 therein), and then using the covariance matrix from the first fit to generate our final best fit model.

\begin{figure}
\includegraphics[width=84mm]{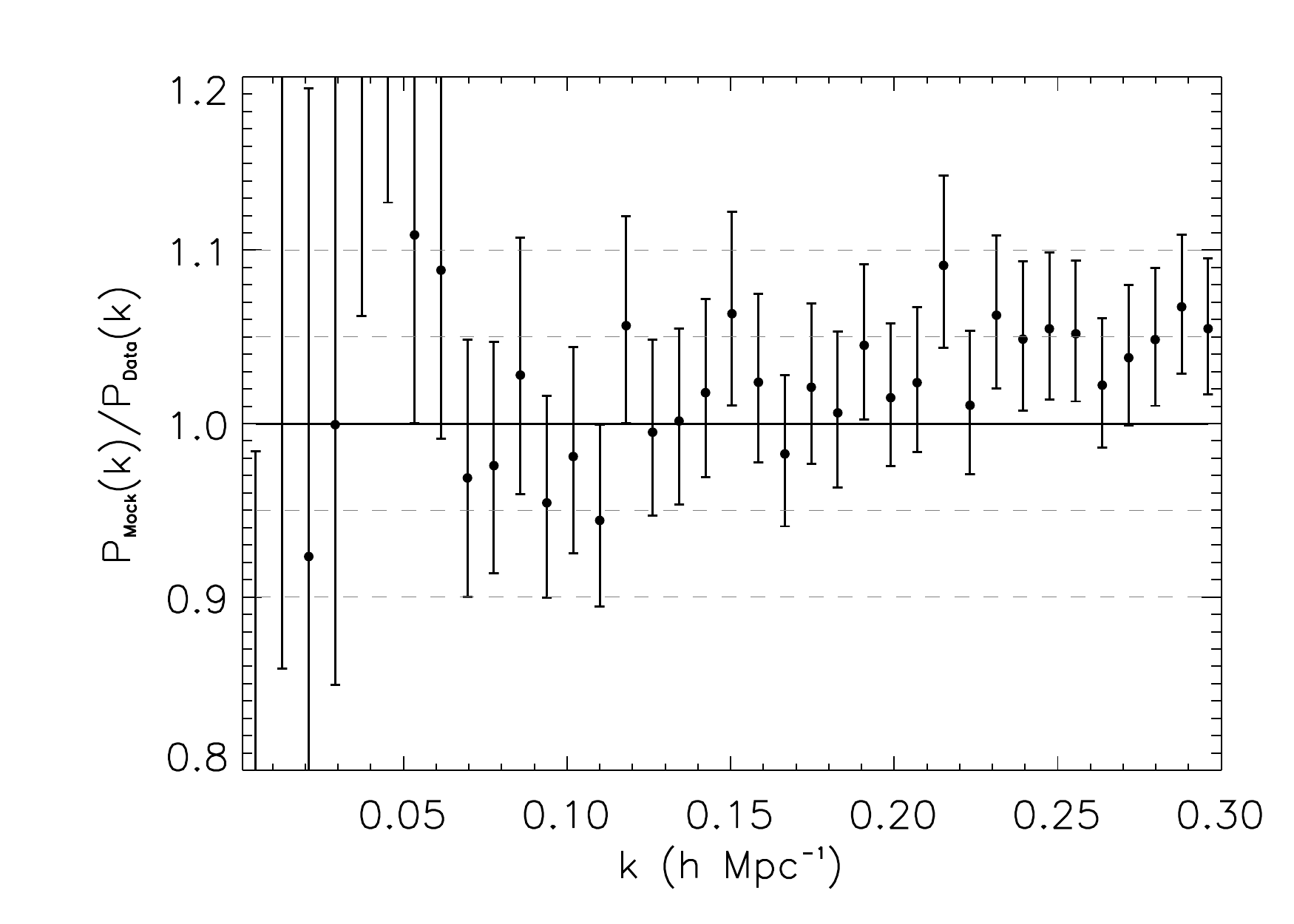}
  \caption{The percentage difference between the average mock power spectrum and that of our data, with errors derived from the covariance matrix of our 1000 mock catalogues. There is good agreement ($\sim5\%$) between these up to $k=0.3$ except on large scales (small $k$) where the window function introduces additional covariance between different k-bins.}
  \label{pkmockdiff}
\end{figure}

Our best fit HOD model has the parameters
\begin{align}
 &\text{log}_{10}(M_{min}) = 13.18, \notag \\
 &\text{log}_{10}(M_{cut})  = 13.15, \notag \\
 &\text{log}_{10}(M_{1}) = 13.94, \notag \\
 &\sigma_{logM} = 0.904, \notag \\
 &\alpha = 1.18, \notag \\
 & \bar{n} = 7\times10^{-4}\,h^{3}\,{\rm Mpc}^{-3}, \notag
\end{align}
where $\bar{n}$ is dependent on the five other parameters. The best fit HOD parameters are in good agreement with the HOD parameters reported by \cite{Zeh11} for another SDSS galaxy sample with similar number density and magnitude limit. Fig.~\ref{pkmockdiff} shows the percentage difference between the average mock power spectrum and the power spectrum of the data. The errors come from the covariance estimated from the full mock sample. We can see that the amplitude of the power spectra matches well on all scales, with $\sim 5\%$ agreement up to $k = 0.3$, except on the largest scales where the window function has a large effect. The fit is good, as we find $\chi^{2} = 33$ for $32$ degrees of freedom (37 $k$-bins and 5 free parameters).

\begin{figure}
\includegraphics[width=84mm]{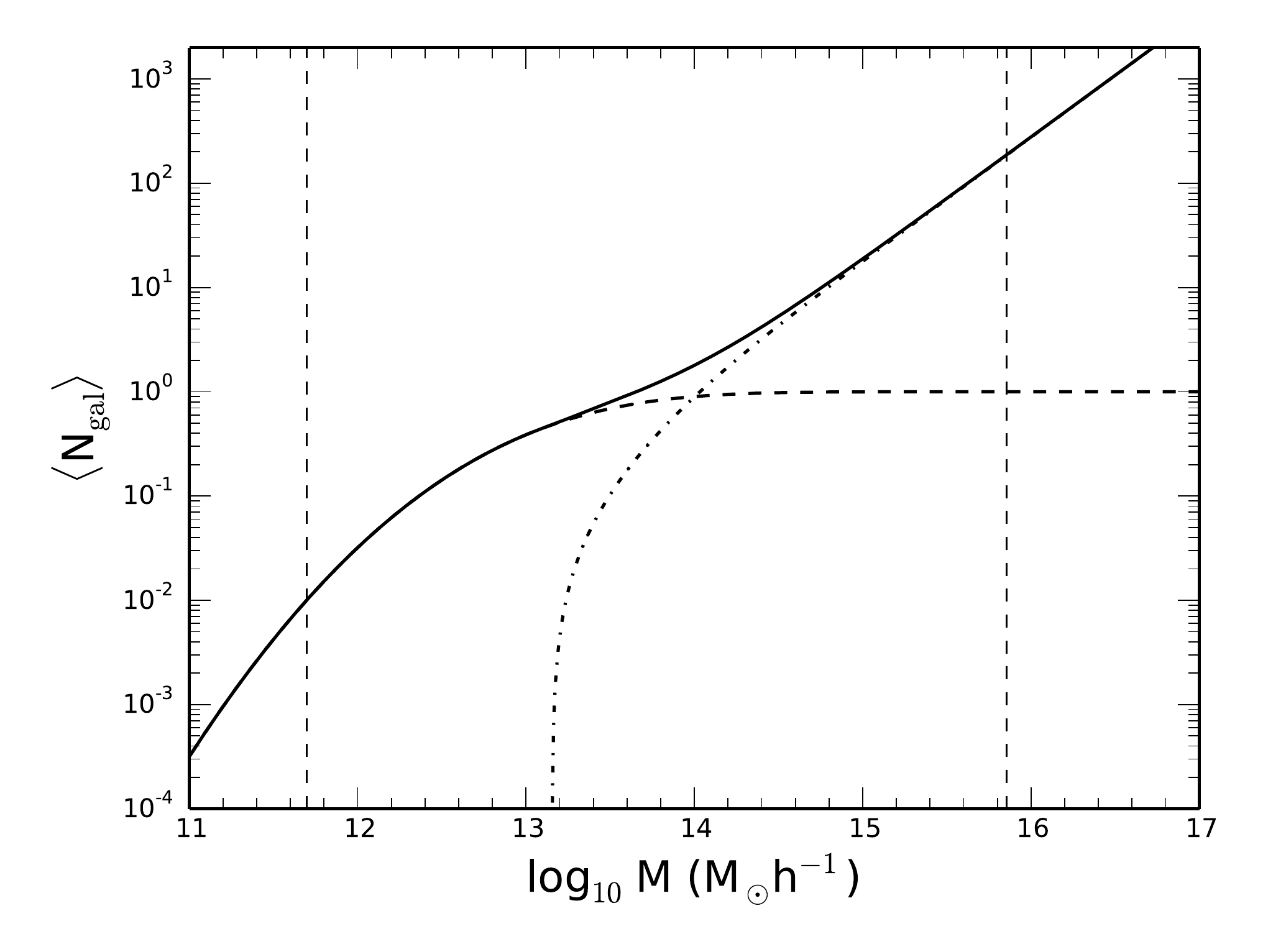}
  \caption{The expected number of galaxies in a halo as a function of halo mass for our bestfit HOD parameters. The dashed line shows the probability of the halo hosting a central galaxy, and the dot-dashed line shows the average number of satellite galaxies within such a halo. The two vertical dashed lines denote the maximum and minimum halo masses across all 1000 mock catalogues.}
  \label{hodnumber}
\end{figure}

Fig.~\ref{hodnumber} shows the expected number of galaxies in our mock halos for our best fit HOD model. This highlights how we are able to recover the clustering properties of the data even though we lack the correct number of low mass halos. All of the satellite galaxies exist in halos with $M > 10^{13} h^{-1}\rm{M_{\odot}}$, which are recovered quite well by our simulations. Below this mass, where our simulations lack sufficient number density, the probability of finding any galaxies within a halo also drops rapidly, such that even though these halos are more abundant in general, the contribution to the total clustering from these halos is small in comparison to the larger mass halos. 

There exists significant degeneracy between the five free HOD parameters, which cannot be broken completely by just the one-dimensional, two-point clustering statistics. Three-point statistics could be used to break this degeneracy \citep{Kulkarni2007}, however this would be prohibitively time-consuming and potentially very noise dominated. Another possibility is to use the quadrupole or hexadecapole moments of the power spectrum, as these contain additional information about the position and velocity distribution of the satellite galaxies within their host halos \citep{Hikage2014}. Again, however, in our case these statistics will almost certainly be noise dominated, and are consequently not important for our current implementation of the method. As such we leave these as future improvements for our mock catalogue production process.

\subsection{Correlation Function}

We base our cosmological fits on configuration-space clustering measurements, calculating the correlation function for both mocks and data as a function of both the redshift space separation $s$, and the cosine of the angle to the line of sight $\mu$, using the same coordinate transformation as for the power spectrum. We use the minimum variance estimator of \cite{LS}, with galaxy and random weights as given in Eq. (\ref{weights}), to calculate the correlation function from the normalised galaxy-galaxy, galaxy-random and random-random pair counts for $0 < s \le 200$ and $0 \le \mu \le 1$ in bins of $\Delta s = 1.0\mpcoh$ and $\Delta \mu = 0.01$.

From there we perform a multipole expansion of the two-dimensional correlation function via the Riemann sum
\begin{equation}
\frac{2\xi_{\ell}(s)}{2\ell+1} = \sum^{100}_{i=1} 0.01\xi(s,\mu_i)P_{\ell}(\mu_i), 
\label{moments}
\end{equation}
where $\mu_i = 0.01i-0.005$ and $P_{\ell}(\mu)$ are the Legendre Polynomials of order $\ell$. We generate the monopole and quadrupole for different bin widths by re-summing the pair counts before applying Eq. (\ref{moments}). 

\begin{figure}
\includegraphics[width=84mm]{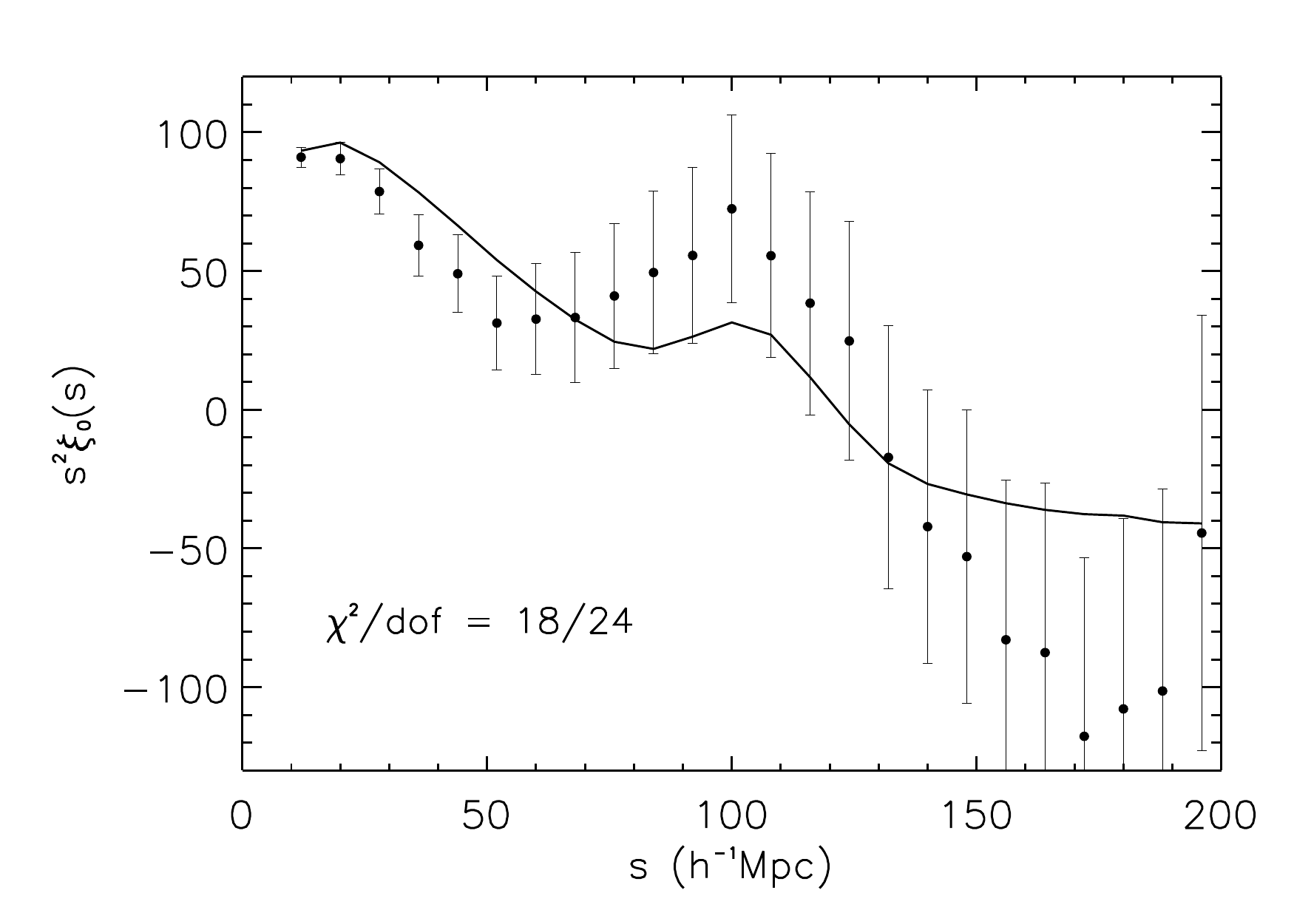}
  \caption{The monopole moment of the correlation function of the MGS. The solid line shows the mean of the mocks and the error bars come from the diagonal elements of the covariance matrix calculated from our 1000 mock realisations.}
  \label{xi0mock}
\end{figure}

\begin{figure}
\includegraphics[width=84mm]{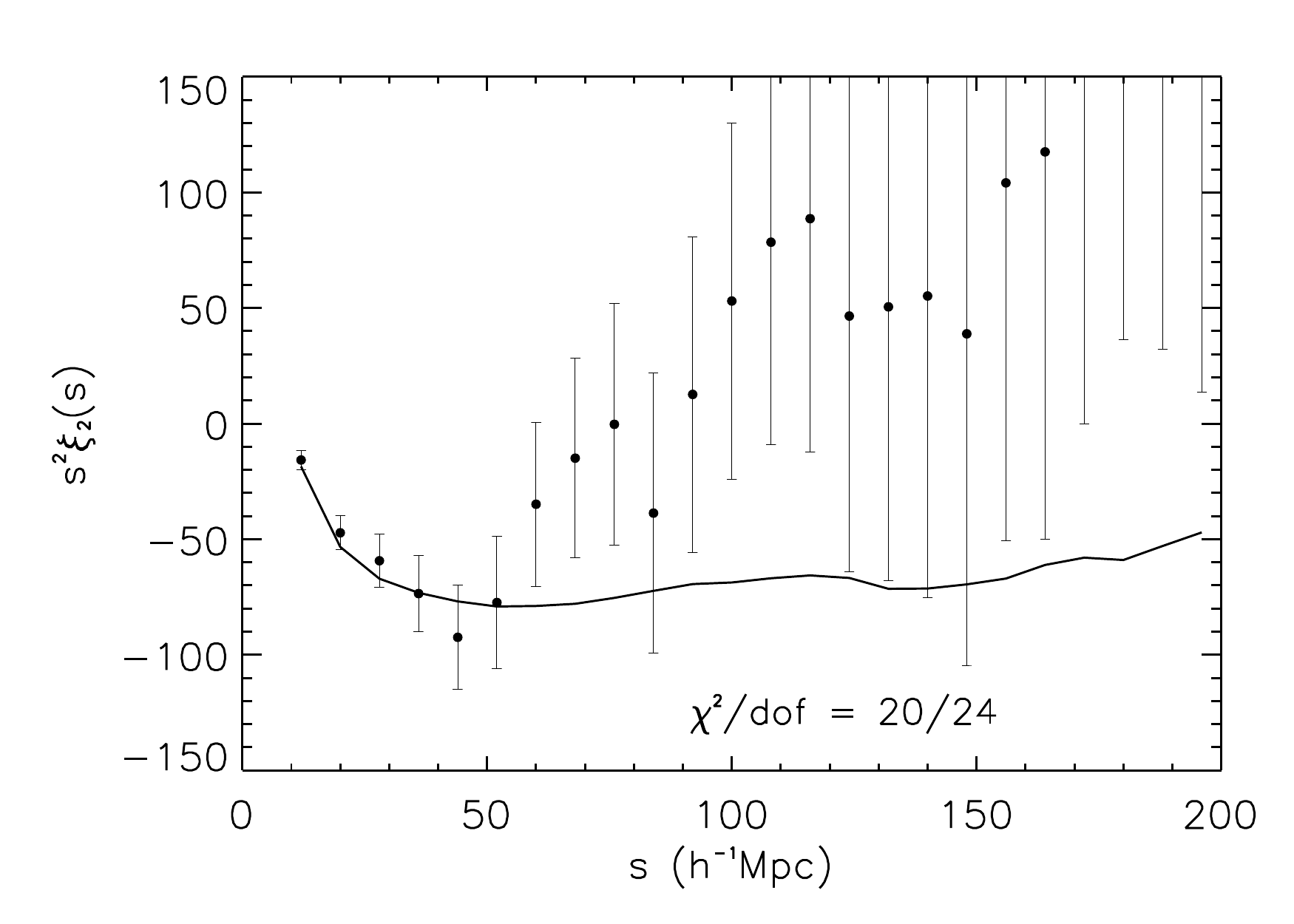}
  \caption{The quadrupole moment of the correlation function of the MGS and the mean of our mock galaxy catalogues. Though the agreement by eye looks poor on large scales, there exists significant covariance between the points at different scales, such that the chi-squared between the data and mocks is small.}
  \label{xi2mock}
\end{figure}

\begin{figure*}
\centering
\subfloat{\includegraphics[trim = 10mm 10mm 15mm 10mm, width=0.32\textwidth]{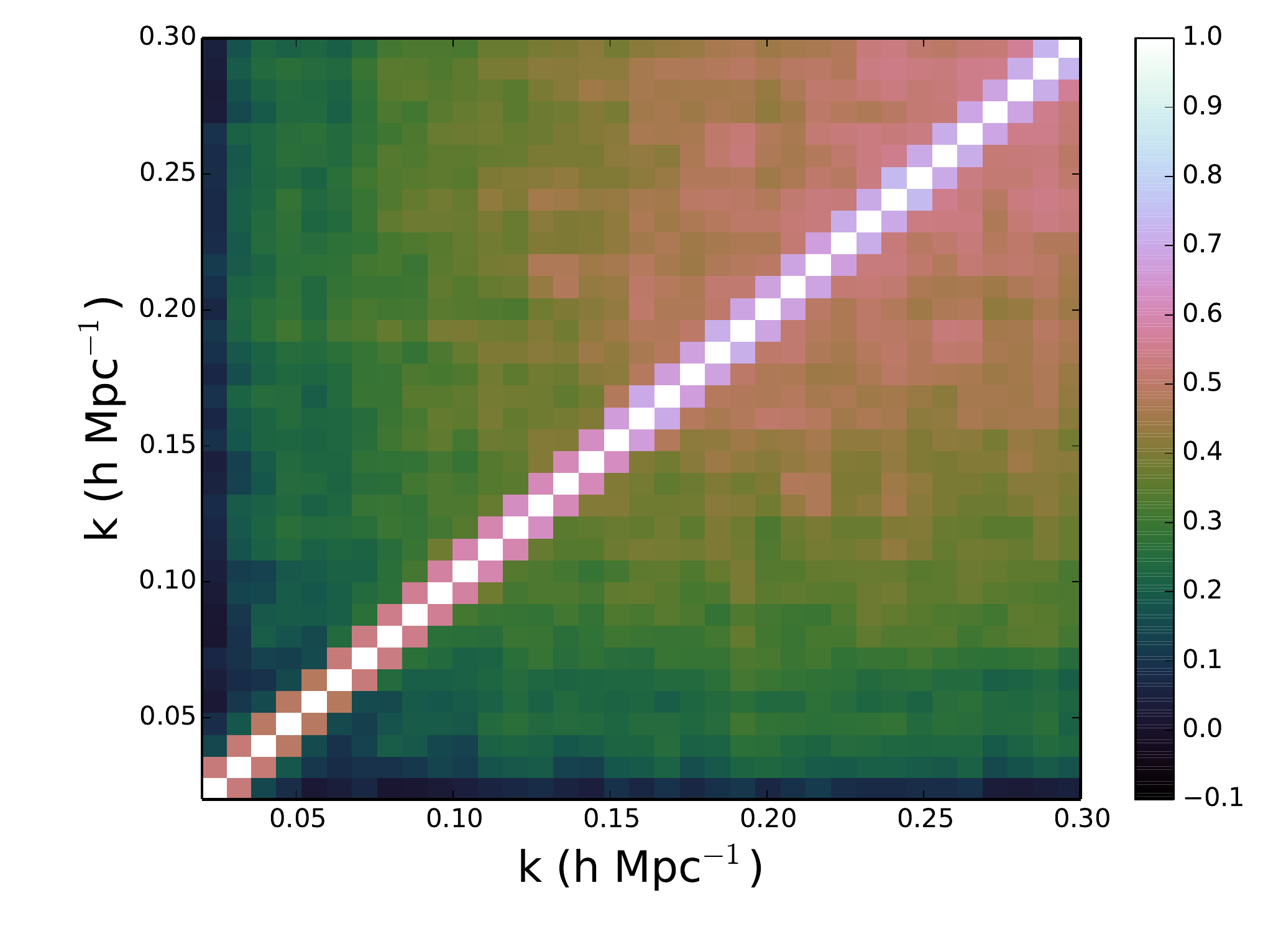}} \\
\subfloat{\includegraphics[trim = 10mm 10mm 15mm 10mm, width=0.32\textwidth]{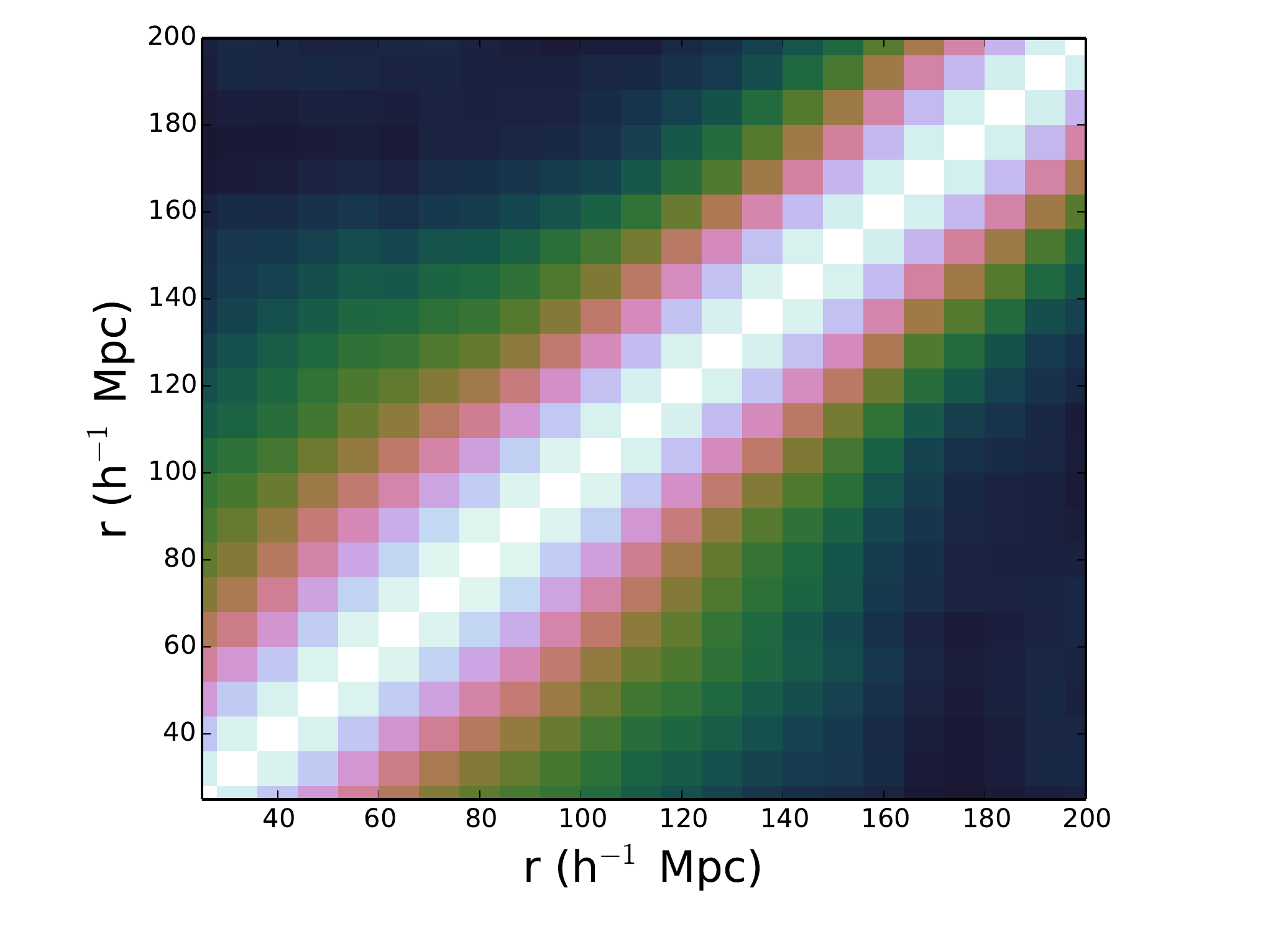}}
\subfloat{\includegraphics[trim = 10mm 10mm 15mm 10mm, width=0.32\textwidth]{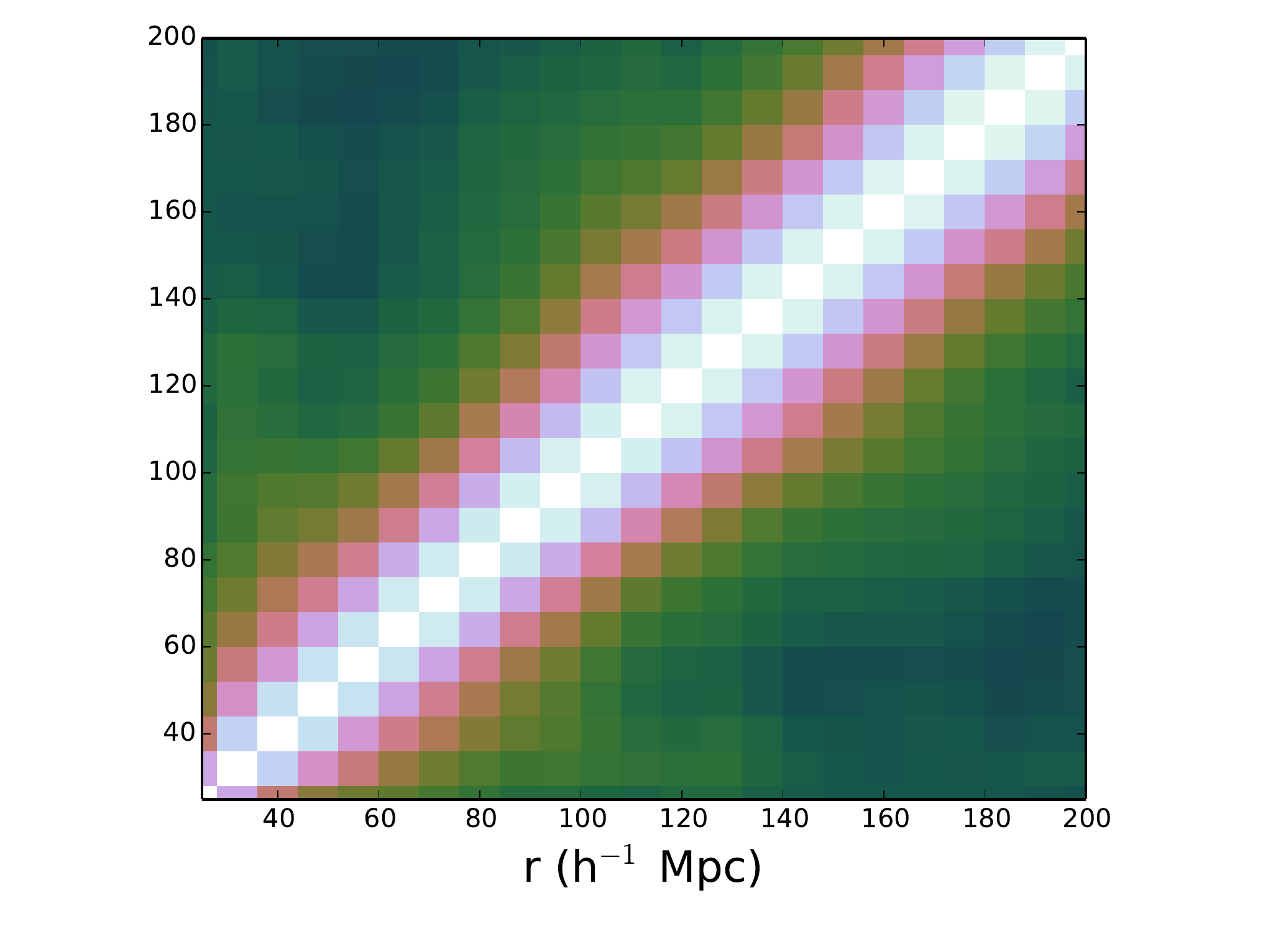}}
\subfloat{\includegraphics[trim = 10mm 10mm 15mm 10mm, width=0.32\textwidth]{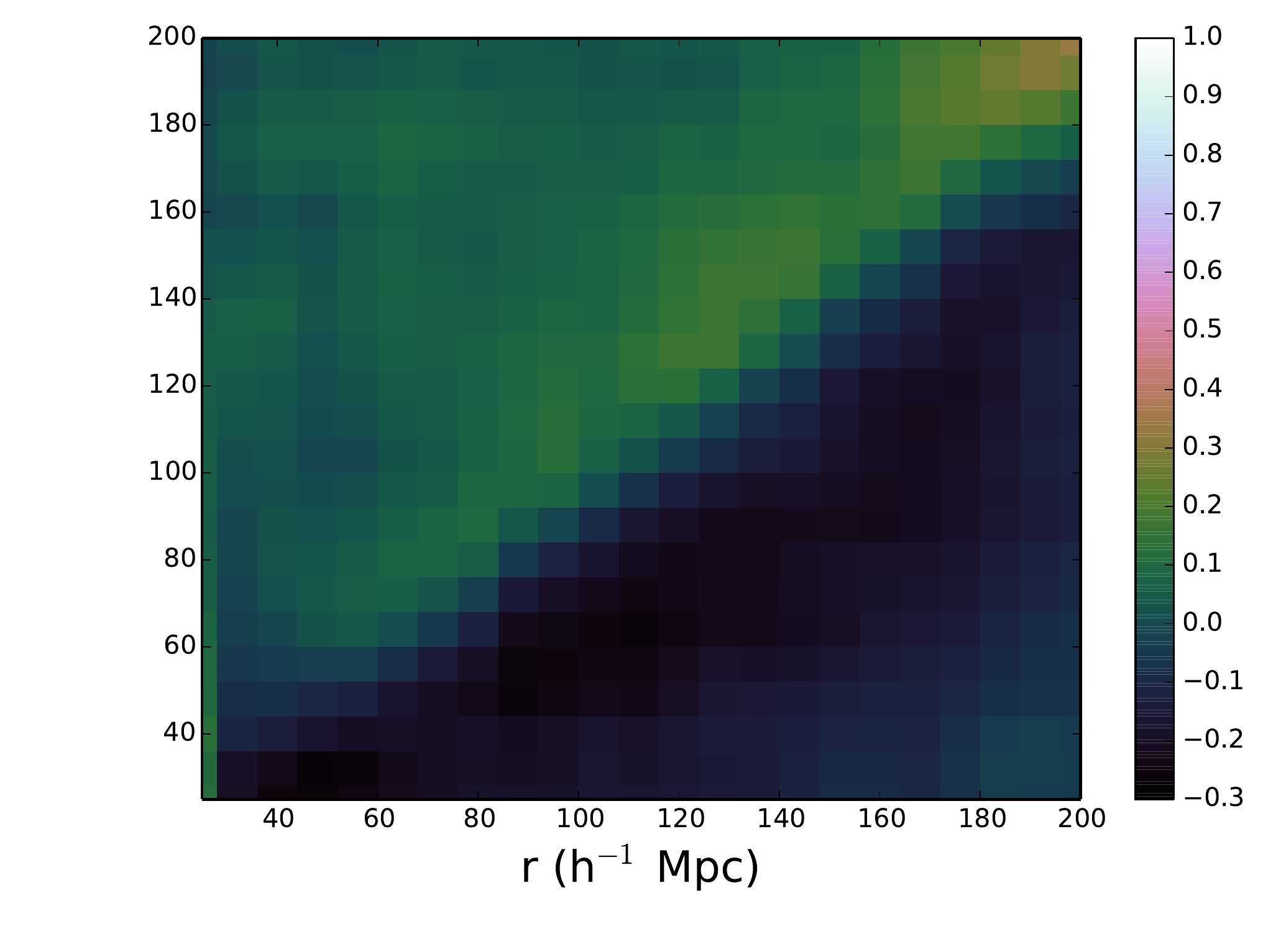}}
\caption{\textit{Top:} The power spectrum correlation matrix generated from our 1000 mock catalogues between $k=0.02\hompc$ and $k=0.3\hompc$ and in bins of $\Delta k = 0.008$. \textit{Bottom:} The correlation matrix for the correlation function monopole (left) and quadrupole (middle) and the cross covariance between the two (right), in bins of $8\mpcoh$ in the range $25 \mpcoh \le s \le 200 \mpcoh$ .}
\label{covar}
\end{figure*}

Figs. \ref{xi0mock} and \ref{xi2mock} show the monopole and quadrupole of the correlation function for the average of the mocks and for the data for the 24 measurements in the range $8 < s < 200 h^{-1}$Mpc. The mean of the mock $\xi_0$ and $\xi_2$ do not match the data within the error-bars at many scales. However, we only plot the diagonal elements of the covariance matrix and the off-diagonal elements represent a significant component (see Fig. \ref{covar}). A more proper comparison is the $\chi^2$ between the mean of the mocks and the data, using the full covariance matrix. For both $\xi_0$ and $\xi_2$ the $\chi^2$/d.o.f is slightly less than one, implying the anisotropic clustering in the mock samples is a good representation of the data, even at 10$h^{-1}$Mpc scales (and hence `$\chi$ by eye' is a bad idea).

\subsection{Covariance Matrix}

We use our sample of mock galaxy catalogues to estimate the covariance matrix for both the power spectrum and correlation function in the standard way, and invert to give an estimate of the inverse matrix. We remove the bias in the inverse covariance matrix by rescaling by a factor that depends on the number of mocks and measurement bins (e.g. \citealt{Hartlap07}).

Fig.~\ref{covar} shows the correlation matrix, $C^{red}_{i,j} = C_{i,j}/\sqrt{C_{i,i}C_{j,j}}$, for the power spectrum and the monopole and quadrupole moments of the correlation function using our fiducial binning scheme. We can see that there is significant off-diagonal covariance in the correlation function and non-negligible cross-covariance between the monopole and quadrupole, however the power spectrum covariance matrix is much more diagonal.

To fit to the correlation function moments, we assume that the binned monopole and quadrupole are drawn from a multi-variate Gaussian distribution, and assume the standard Gaussian Likelihood, $\mathcal{L}$. The validity of this assumption, for both our fits and the BAO fits to the power spectrum found in Paper I, is tested in the following section. There are additional factors that one must apply to uncertainties determined using a covariance matrix that is constructed from a finite number of realisations and to standard deviations determined from those realisations \citep{DS12,Per14}. In this work we multiply the inverse covariance matrix estimate by a further factor given by $m_1$ in equation~18 of \citet{Per14}, such that the errors derived from the shape of the likelihood are automatically corrected for this bias. We have the number of mocks $N_{\rm mocks}=1000$, the number of bins $N_{\rm bins}=34$ and the number of parameters fitted $N_{\rm p}=8$, giving only a small correction to the inverse covariance matrix of $1.02$.

\subsection{Systematic Tests}

\subsubsection{Independence of mocks}

\begin{figure}
\includegraphics[width=84mm]{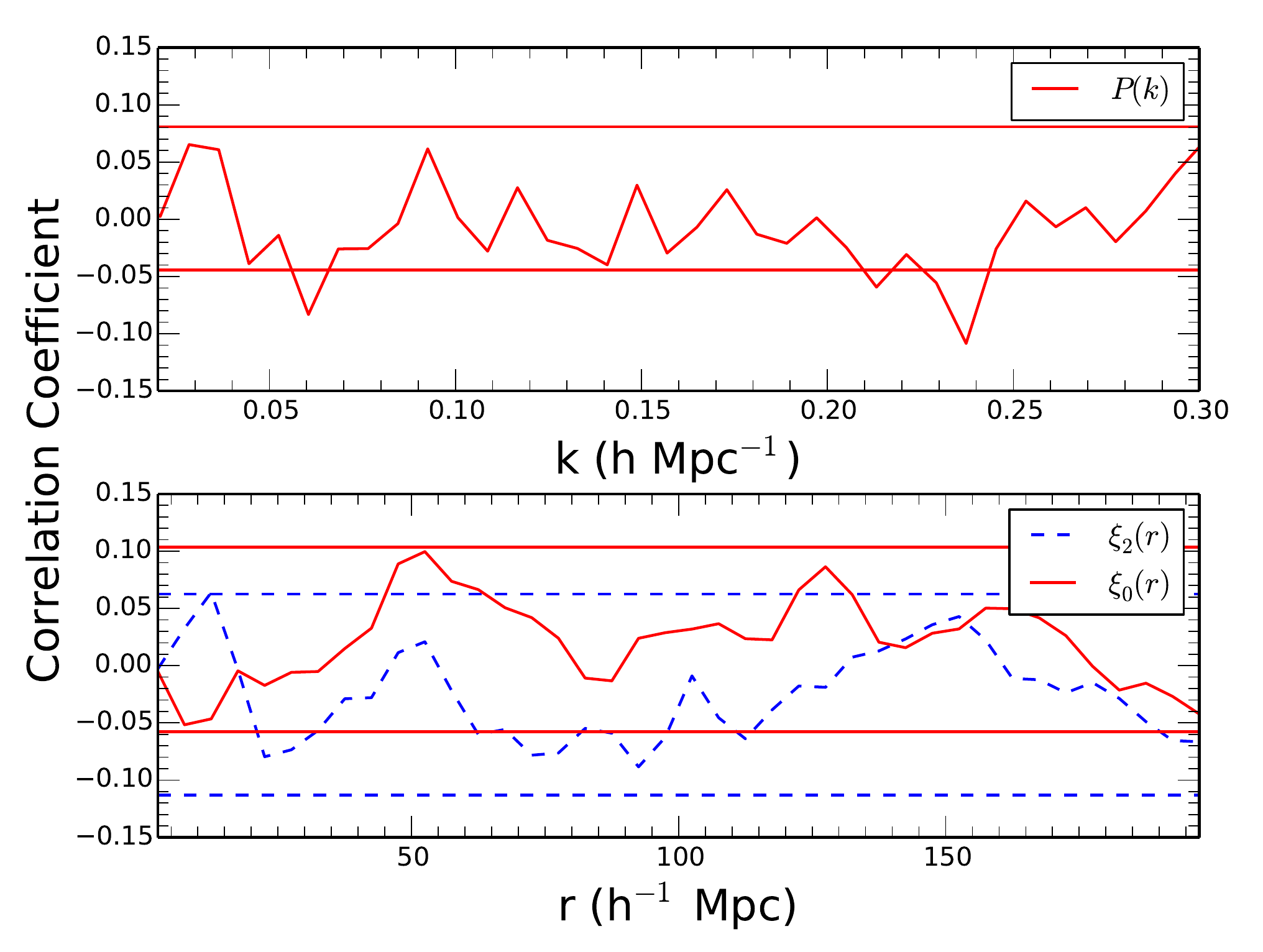}
  \caption{The cross-correlation coefficient between pairs of mocks generated from the same dark matter field, for both the power spectrum and the monopole and quadrupole of the correlation function. The horizontal lines indicate the maximum and minimum (across all scales) cross-correlation measured from an equivalent number of pairs of mocks that are drawn from different dark matter realisations.}
  \label{crosscoeff}
\end{figure}

The coordinate transformation that allows us to create two distinct mocks from each dark matter realisation puts the two patches as far apart as possible to minimise the covariance between mocks based on the same dark matter cube. The minimum possible distance between two objects in different patches is $170\mpcoh$. Whilst this is within the range of scales we are interested in, the total cross-correlation between patches is very small. We number our mocks such that pairs of mocks (e.g. 1 \& 2, or 3 \& 4) were drawn from the same dark matter cube. Thus we expect the set of $500$ even numbered mocks and the set of $500$ odd numbered mocks to be independent of any correlations caused by the sampling, and any cross correlation to be due to noise. The cross correlation coefficient, 
\begin{equation}
\rho_{X,Y} = \frac{\textbf{\sf C}(X,Y)}{\sigma_{X}\sigma_{Y}}
\end{equation}
for both the monopole and quadrupole of the correlation function, and for the power spectrum, calculated from the 500 pairs of mocks drawn from the same dark matter cube is shown in Fig.~\ref{crosscoeff}. The dashed lines in Fig.~\ref{crosscoeff} indicate the maximum and minimum correlation coefficient (at any scale considered) between 500 pairs of independent mocks (i.e. taking pairs where both mocks have even or odd numbers). The fact that the cross correlation between pairs drawn from the same dark matter cube is almost entirely within these bounds indicates that there is no cross correlation above the level of noise in our combined covariance matrix, even on scales where the pairs of mocks could, theoretically, be covariant.

\subsubsection{Random catalogue redshift assignment}
 
\begin{figure}
\includegraphics[width=84mm]{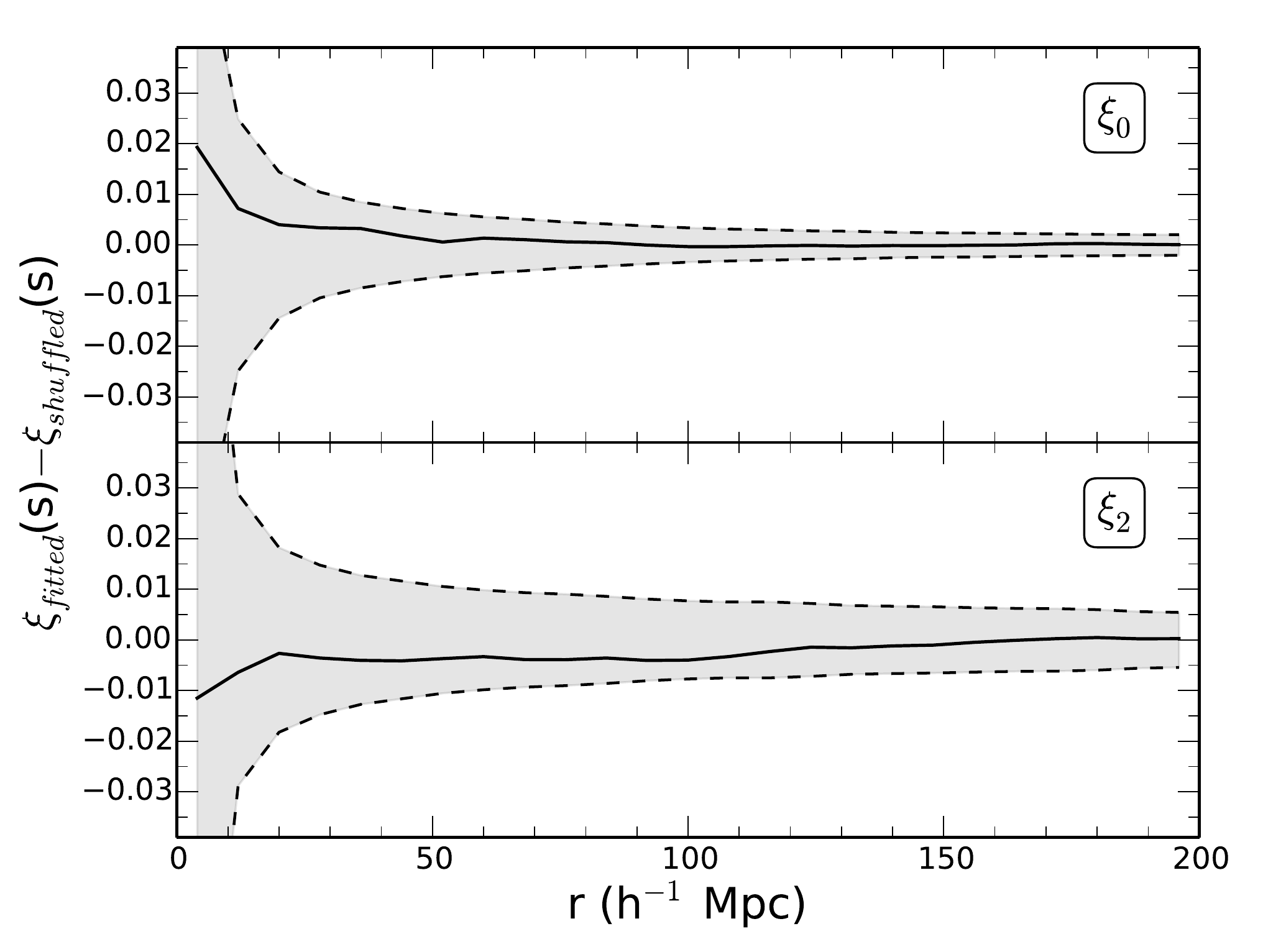}
  \caption{The difference in the monopole and quadrupole of the correlation function measured from the data when we use the fitted and shuffled methods of generating redshifts for our random data points. The shaded areas denote the one-sigma error regions. We see that the difference between the two methods is well within the one-sigma region on all scales.}
  \label{reshuffled}
\end{figure}

We also test the effect of assigning redshifts to our random data points from randomly chosen galaxies as opposed to simply generating them by sampling a smooth fit to the number density. In Fig.~\ref{reshuffled} we present the differences in the measured correlation function monopole and quadrupole moments of the MGS data, when they are calculated either using random data points that are assigned redshifts from the corresponding galaxy catalogue (`shuffled'), or when they are given redshifts sampled from the fitted number density described in Section~\ref{sec:data}. We may expect `shuffling' to reduce the clustering, especially on scales below $100\mpcoh$, because spherically averaged features in the galaxy field are removed in the shuffled approach. The power removed is predominantly along the line of sight, and hence the quadrupole is affected more than the monopole. From Fig.~\ref{reshuffled} we see that for both monopole and quadrupole, the difference in clustering between the two methods is well below the level of the noise. We adopt the shuffling approach as we do not know the true radial distribution for the data, and this approach allows for all features caused by the galaxy selection, at the expense of a small reduction in the monopole and quadrupole moments. Further, \cite{DR9sys} found that the shuffling approach is less biased than fitting to a smooth $n(z)$ when both methods were tested on BOSS mocks (with a known $n(z)$), and the differences we find are consistent with those of \cite{DR9sys}. Such differences are so small that we do not need to account for this in our model fitting.

\subsubsection{Gaussianity of data}
 
\begin{figure}
\includegraphics[width=84mm]{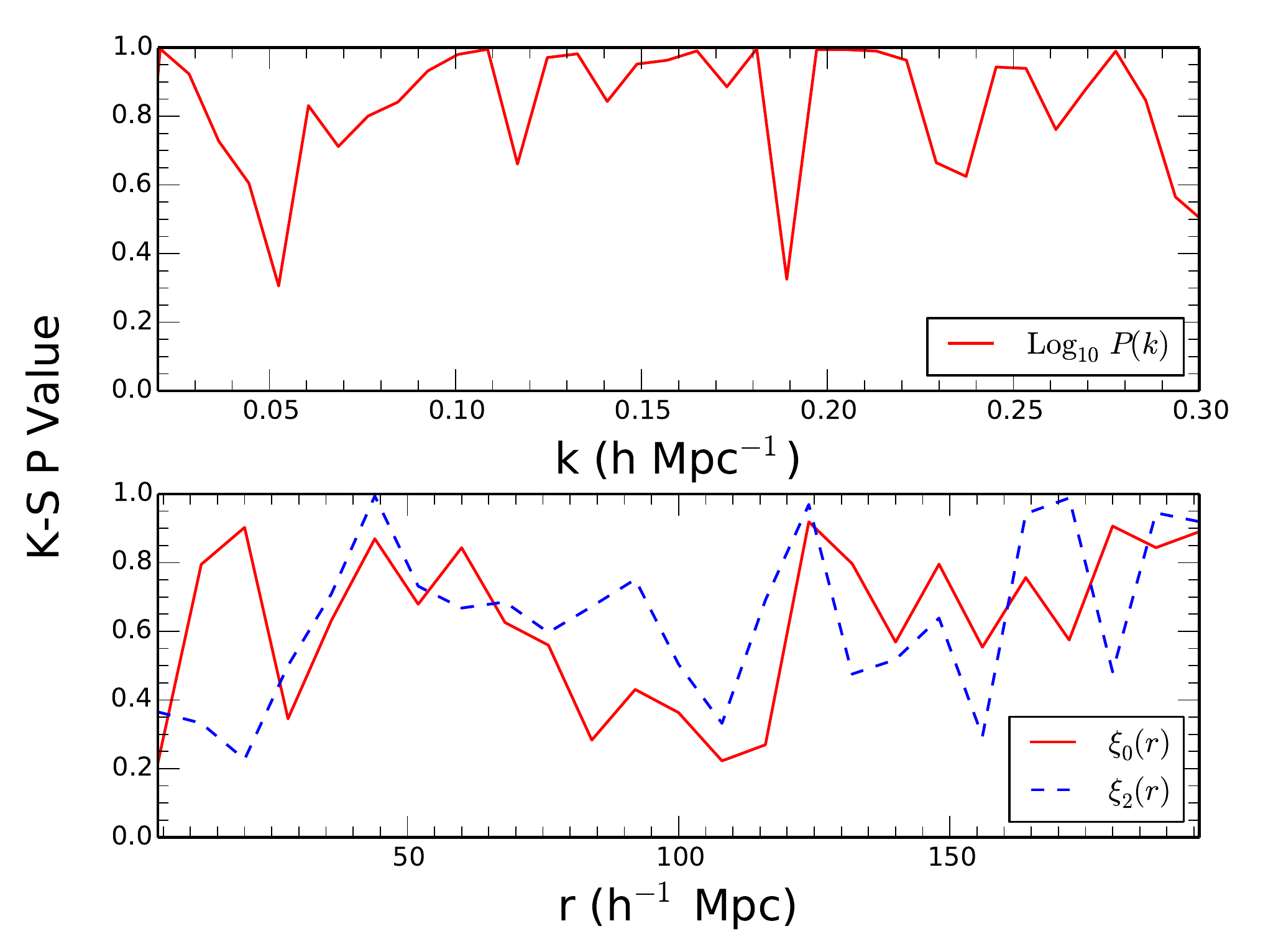}
  \caption{The Kolmogorov\hyp{}Smirnov p-value for both the log of the power spectrum and the monopole and quadrupole of the correlation function. For both statistics the probability that they are drawn from a multivariate Gaussian is high, allowing us the compute the likelihoods for theoretical models from the chi-squared difference between the model and data.}
  \label{kstest}
\end{figure}

Our final test is on the assumption that the measured correlation function and power spectrum are drawn from an underlying multivariate Gaussian distribution. This assumption is the basis of the likelihood calculations made in both the BAO fits of Paper I and the RSD fits presented in this paper. 

We perform a Kolmogorov\hyp{}Smirnov test on the log of the power spectrum (which is used in the BAO fits of Paper I) and monopole and quadrupole of our mock catalogues, using the cumulative distribution function (CDF) of the normalised differences between the two-point statistics measured from each mock realisation and the average over all the mock catalogues. 
Following the standard method of the Kolmogorov\hyp{}Smirnov test we define the parameter $D$ as the maximum difference between our CDF and the CDF of the distribution we wish to test against, in this case a Gaussian. The p-value for this test, which indicates the probability that the observed value of $D$ would be a large as it is if our underlying distribution \textit{were} Gaussian, is then given by a simple rescaling of the parameter D,
\begin{equation}
D^{*} = D\left(\sqrt{N} + \frac{0.11}{\sqrt{N}} + 0.12\right),
\end{equation}
and the the approximate expression
\begin{equation}
P(D > D_{obs}) \approx 2\sum_{k=1}^{\infty}(-1)^{k-1}e^{-2k^{2}D^{*}}.
\end{equation}
Here, $N$ is the number of bins in our measured CDF. As elsewhere, we use bins of width $\Delta k = 0.008$ for the power spectrum and $\Delta s = 8\mpcoh$ for the correlation function.

Fig.~\ref{kstest} shows the Kolmogorov\hyp{}Smirnov test p-value for the two-point statistics as a function of scale. We can see that there is no trend with scale and across all scales of interest the p-value indicates a high probability that both the power spectrum and correlation function are drawn from a Gaussian distribution. The log of the power spectrum has a particularly high probability of being drawn from a Gaussian distribution, which is why we use this rather than the power spectrum itself when fitting the BAO feature in Paper I. Based on the p-values we obtain, we find that even for those bins in the correlation function where the difference between our measured CDF and a Gaussian CDF is largest, we could expect a greater difference at least 20\% of the time if our measured clustering statistics were drawn from an underlying Gaussian distribution.

\section {Modelling the redshift space monopole and quadrupole}
\subsection{Modelling the Effect of Galaxy Velocities}
To model our redshift space monopole and quadrupole we use the combined Gaussian Streaming/Convolved Lagrangian Perturbation Theory (CLPT)  model of \cite{Wang2014}. The clustering of galaxies in redshift space can be written as a function of their real space correlation and their full pairwise velocity dispersion \citep{Fisher1995, Scoccimarro2004}. In the Gaussian Streaming model, introduced by \cite{Reid2011}, the pairwise velocity dispersion is approximated as a Gaussian, which allows one to write the two-dimensional redshift space correlation function, $\xi(s_{\perp},s_{||})$, as a function of the real-space correlation function, $\xi(r)$, and the mean infall velocity and velocity dispersions betweens pairs of galaxies, $v_{12}(r)$ and $\sigma_{12}^{2}(r,\mu)$ respectively,
\begin{equation}
\begin{aligned}
1+\xi(s_{\perp},s_{||}) &= \int_{-\infty}^{\infty}\frac{dr_{||}}{[2\pi\sigma_{12}^{2}(r,\mu)]^{1/2}}[1+\xi(r)] \\
&\times \text{exp} \left\{ -\frac{[s_{||}-r_{||}-\mu v_{12}(r)]^{2}}{2\sigma_{12}^{2}(r,\mu)} \right\}.
\label{eq:GS}
\end{aligned}
\end{equation}
Here $s_{\perp}=r_{\perp}$ and $s_{||}$ denote redshift space separations transverse and parallel to the line of sight, $r_{||}$ denotes the real space separation parallel to the line of sight, such that $r^{2}=r_{\perp}^{2}+r_{||}^{2}$, and $\mu = r_{||}/r$ is as defined previously.

\cite{Reid2011} evaluate $v_{12}(r)$ and $\sigma_{12}^{2}(r,\mu)$ using a standard perturbation theory expansion of a linearly biased tracer density field, however this does not accurately replicate the velocity statistics of the tracer field on small scales, nor the smoothing of the BAO feature. This was improved upon by \cite{Reid2012} in their analysis of the BOSS CMASS galaxy sample by using Lagrangian Perturbation Theory to generate the real-space correlation function above scales of $70\mpcoh$. This proved effective for the BOSS CMASS sample, although \cite{Reid2012} note that the BOSS CMASS galaxy sample has a second order bias close to zero, the point at which the accuracy of the standard perturbation theory evaluation of $v_{12}(r)$ and its derivative is greatest. 

\cite{Carlson2013} and \cite{Wang2014} further improve the modelling of the correlation function by computing the real-space correlation function using Convolved Lagrangian Perturbation Theory and evaluating $v_{12}(r)$ and $\sigma_{12}^{2}(r,\mu)$ in the same framework. This formulation relies on a perturbative expansion of the Lagrangian overdensity and displacement which in turn allows us to write the  correlation function and velocity statistics as a series of integrals over powers of the linear power spectrum. For biased tracers the model assumes a local real-space Lagrangian bias function, $F$, and solutions up to $\mathcal{O}(P_{L}^{2})$ reveal a dependence on both the first and second derivatives of the bias function, $\langle F' \rangle$ and $\langle F'' \rangle$, and combinations thereof. Furthermore, as would be expected, the velocity statistics have a dependency on the growth rate of structure, $f$, via the multiplicative factor, $f^{2}$. From \cite{Matsubara2008} we can easily relate the linear galaxy bias, $b$, to the first derivative of the Lagrangian bias function by $\langle F' \rangle = b-1$.

The model is calculated as follows. For a vector $\boldsymbol{r}$ in real space and vector $\boldsymbol{q}$ in Langrangian space, we can define three functions that depend on the Lagrangian bias, growth rate and linear power spectrum: $M_{0}(\boldsymbol{r},\boldsymbol{q},\langle F' \rangle, \langle F'' \rangle, f, P_{L})$, $M_{1,n}(\boldsymbol{r},\boldsymbol{q},\langle F' \rangle, \langle F'' \rangle, f, P_{L})$ and $M_{2,nm}(\boldsymbol{r},\boldsymbol{q},\langle F' \rangle, \langle F'' \rangle, f, P_{L})$. $M_{0}$ is a scalar function, whilst $M_{1,n}$ and $M_{2,nm}$ are vector and tensor functions along cartesian directions $n$ and $m$. The exact form of the functions $M_{0}$, $M_{1,n}$, and $M_{2,nm}$ are given in \cite{Wang2014}

We can then calculate $\xi(r)$ and $v_{12}(r)$ by projecting the scalar and vector functions along the pair separation vector and integrating with respect to the Lagrangian separation,
\begin{align}
1+\xi(r) &= \int d^{3}q M_{0}(\boldsymbol{r},\boldsymbol{q}), \\
v_{12}(r) &= [1+\xi(r)]^{-1}\int d^{3}q M_{1,n}(\boldsymbol{r},\boldsymbol{q})\hat{r}_{n}.
\end{align}
We split the velocity dispersion $\sigma_{12}^{2}(r,\mu)$ into components perpendicular and parallel to the pair separation vector and evaluate these separately by projecting and integrating the tensor function,
\begin{equation}
\sigma_{12}^{2}(r, \mu) = \mu^{2}\sigma_{||}^{2}(r) + (1-\mu^{2})\sigma_{\perp}^{2}(r),
\end{equation}
where
\begin{align}
\sigma_{||}^{2}(r) &= [1+\xi(r)]^{-1}\int d^{3}q M_{2,nm}(\boldsymbol{r},\boldsymbol{q})\hat{r}_{n}\hat{r}_{m}, \\
\sigma_{\perp}^{2}(r) &= \frac{[1+\xi(r)]^{-1}}{2}\int d^{3}q M_{2,nm}(\boldsymbol{r},\boldsymbol{q})\delta_{nm}^{K} - \frac{\sigma_{||}^{2}}{2}
\end{align}
and $\delta_{nm}^{K}$ is the Kronecker delta.

Hence, for a given cosmological model parameterised by $P_{L}, b, \langle F'' \rangle$ and $f$, we can calculate, for any scale of interest, a unique set of $\xi(r)$, $v_{12}(r)$ and $\sigma_{12}^{2}(r,\mu)$. Entering these into Eq. (\ref{eq:GS}) allows us to generate our two-dimensional redshift space correlation function and from there we can generate a model monopole and quadrupole. These models are fitted to the measurements from data and mocks as described later to constrain a given set of cosmological parameters.

\subsection{Alcock-Paczynski Effect}

In calculating the correlation function of our data we have to assume a (fiducial) cosmological model to calculate the physical separations between galaxies parallel and transverse to the line of sight. Specifically, to calculate the separation along the line of sight we require the Hubble parameter, $H(z)$, and the galaxy redshifts, whilst the transverse separation requires knowledge of the angular diameter distance, $D_{A}(z)$, and the angular separation of the galaxy pair. Any difference between the relative values of these parameters in the fiducial cosmology and the true cosmology will manifest as anisotropic clustering, that is, a difference in the clustering of galaxies parallel and perpendicular to the line of sight. If an observable such as the BAO feature is expected to be statistically isotropic, then any measured anisotropy can also be used to constrain the true cosmology of our universe. This is the Alcock-Paczynski (AP) test \citep{AP}.
 
Anisotropy is also being added by Redshift Space Distortions. As such, the AP effect and RSD are degenerate and we need a way to disentangle these effects. 

Following \cite{Xu2013}, we introduce two scale parameters, $\alpha$ and $\epsilon$. $\alpha$ denotes the stretching of all scales and hence encapsulates the isotropic shift whilst $\epsilon$ parameterises the AP effect.  Measuring these two parameters allows us to constrain the angular diameter distance and Hubble expansion independently, 
\begin{align}
\alpha &=  \left( \frac{D_{A}^{2}(z)}{D_{A,fid}^{2}(z)} \frac{H_{fid}(z)}{H(z)} \right)^{1/3} \frac{r_{s,fid}}{r_{s}}, \\
1+\epsilon &= \,\,\, \frac{F(z)}{F_{fid}(z)} \,\,\,= \left( \frac{D_{A,fid}(z)}{D_{A}(z)}  \frac{H_{fid}(z)}{H(z)} \right)^{1/3}.
\end{align}
where a subscript `fid' denotes our fiducial model and $r_{s}$ is the measured BAO peak position. Values $\alpha=1.0$ and $\epsilon=0.0$ would indicate that our fiducial cosmology is the true cosmology of the measured correlation function.

In terms of our model correlation function the $\alpha$ and $\epsilon$ parameters modify the scales at which we measure a given value for the correlation function,
\begin{align}
s'_{||} &= \alpha(1+\epsilon)^{2}s_{||}, \notag \\
s'_{\perp} &= \alpha(1+\epsilon)^{-1}s_{\perp}. 
\end{align}

During our fits we apply the values of $\alpha$ and $\epsilon$ directly to alter the scales at which we calculate the two-dimensional redshift space correlation function (given by Eq. (\ref{eq:GS})), calculating the necessary correction to the parallel and perpendicular separations, $s_{||}$ and $s_{\perp}$, before using these to calculate the corresponding values of $r, r_{||}$ and $\mu$ required by the integrand. We subsequently integrate the 2D model for the correlation function to estimate monopole and quadrupole moments.

\subsection{Correction for binning effects}

Finally, we must account for the way we bin our data when calculating our model. Rather than evaluating our model at the centre of those bins, we take into account variations in the model correlation function across each bin, and instead take the weighted average of our model within each bin. For a bin from $s_{1}$ to $s_{2}$ centred at $s$, our model is
\begin{align}
\xi_{0, {\rm mod}}(s) &= \frac{1}{V}\int_{s_{1}}^{s_{2}}\xi_{0}(s')s'^{2}ds',\notag \\
\xi_{2, {\rm mod}}(s) &= \frac{1}{V}\int_{s_{1}}^{s_{2}}\xi_{2}(s')s'^{2}ds'.
\label{eq:bincentre}
\end{align}
Where $V$ is the normalisation for the weighted mean,
\begin{equation}
V=\int_{s_{1}}^{s_{2}} s'^{2} ds'.
\end{equation}

For all the fits detailed in this paper we calculate our model in bins of width $\Delta s = 1\mpcoh$ between $0\mpcoh < s \le 200\mpcoh$, before calculating Eq. (\ref{eq:bincentre}), using a cubic spline interpolation method to interpolate the value of the monopole and quadrupole at any point required for the integration.

\section{Analysis}

\subsection{Cosmological Parameters}

For our analysis, we consider the shape of the linear power spectrum to be parameterised by the cold dark matter and baryonic matter densities, $\Omega_{c}h^{2}$ and $\Omega_{b}h^{2}$, and the scalar index, $n_{s}$, whilst the amplitude of the power spectrum is quantified using $\sigma_{8}$. On top of this we add the growth rate of structure, $f$, which we wish to measure via the RSD signal, galaxy bias parameters $b$ and $\langle F'' \rangle$, and BAO dilation parameters $\alpha$ and $\epsilon$, which we measure independently of the power spectrum shape.

In theory, the dependence of the CLPT model on $P_{L}$, $b$, $f$, $\langle F'' \rangle$ and $\sigma_{8}$ is such that, combined with the other dependencies, all of the above parameters can be independently measured if the data has no noise. In practice however, the parameters $f$, $b$ and $\sigma_{8}$ are strongly degenerate at the linear level and we are unable to constrain these independently. In addition, we can provide no constraints on the shape of the linear power spectrum beyond those, already tight, constraints given by the Planck Collaboration's analysis of the Cosmic Microwave Background radiation. In lieu of this we fix $\Omega_{c}h^{2}$, $\Omega_{b}h^{2}$ and $n_{s}$ to the fiducial values used to create our mock catalogues, which correspond closely to the Planck best-fit values, and assume that any variation in these parameters can be captured by departures from $\alpha=1.00$  and $\epsilon=0.00$.

Overall, then, we explore a combination of cosmological parameters $\vec{p} = \{b\sigma_{8}, \langle F'' \rangle, f\sigma_{8}, \sigma_{8,nl}, \alpha, \epsilon\}$. Here we treat $\sigma_{8}$ as containing two separate contributions, linear and non-linear. The former of these is contained in the $b\sigma_{8}$ and $f\sigma_{8}$ parameters which are our parameters of interest and are responsible for the overall amplitude of the monopole and quadrupole of the correlation function. The latter, $\sigma_{8,nl}$, is only effective at the smallest scales we fit against and as such is largely unconstrained and degenerate with the second order bias parameter $\langle F'' \rangle$.

In all fits we do not allow $f\sigma_{8}$ to vary in such way that we choose unphysical values of $f\sigma_{8} < 0$ or $\sigma_{8,nl} < 0 \,h^{3}\,{\rm Mpc}^{-3}$, and we apply uniform priors of $0.8<\alpha<1.2$ and $-0.2<\epsilon<0.2$, as for the BAO fits of Paper I. We include priors on $\alpha$ and $\sigma_{8,nl}$ as described and tested in Section~\ref{sec:priors}.

\subsection{Nuisance Parameters}

We marginalise over two nuisance parameters while fitting the correlation function, which we denote $\sigma_{offset}$ and $IC$. The first of these corresponds to an additive correction to $\sigma_{12}$ in the Gaussian Streaming model. This compensates for two different effects that both manifest at the same point in the model. The first is the CLPT model's inability to fully recover the large scale halo velocity dispersion. Whilst the scale-dependence of both the $\sigma_{||}$ and $\sigma_{\perp}$ parts of $\sigma_{12}$ is well recovered by the CLPT, there is a mass-dependent, constant amplitude shift across all scales. This systematic offset in the halo velocity dispersion offset is identified in \cite{Reid2011} and further explored in \cite{Wang2014}, who go on to suggest that it stems from gravitational evolution on the smallest scales, which cannot be accurately predicted by perturbation theory and hence cannot be separated from the overall scale-dependence of $\sigma_{12}$. Rather than calibrate the corrective factor required to shift the amplitude of the velocity dispersion using, for example, N-Body simulations we simply treat this as a free parameter, and part of the $\sigma_{offset}$ nuisance parameter. The second component of $\sigma_{offset}$ is the additional velocity dispersion along the line of sight due to the so called, `Fingers-of-God', resulting from peculiar motions of the galaxies within their host halos. This effect is expected to be small on our scales of interest and in the monopole and quadrupole of the correlation function. 

We apply a  very broad, flat prior of $-40\, {\rm Mpc}^{2}<\sigma_{offset} < 40\, {\rm Mpc}^{2}$. This range is similar to that used in \cite{Reid2012}, where they allow the Fingers-of-God intra-halo velocity dispersion to vary from $0 \, {\rm Mpc}^{2}$ to $40 \, {\rm Mpc}^{2}$, providing a detailed set of tests to validate this prior. We additionally allow this term to go negative over the same range to account for the fact that, as mentioned in \cite{Reid2011}, the perturbation theory calculation of $\sigma_{12}$ overestimates the amplitude of the positive offset required to bring linear theory in line with the measurements from N-Body simulations, hence resulting in a $\sigma_{12}$ which is larger than would be measured. 

Our second nuisance term is the integral constraint, which takes the form of an additional constant added to the correlation function monopole. This accounts for incorrect clustering on the largest scales due to the finite volume of our survey. Whilst, given a model, this can be calculated analytically from the properties of our survey, we include it as a free parameter to also account for additional uncertainties in the modelling of the monopole and potential observational systematic effects, which tend to add nearly scale-independent clustering \citep{DR9sys}. Under the assumption that the integral constraint is independent of the angle to the LOS, this vanishes for the quadrupole and so we only apply a nuisance parameter of this form to the monopole. 

\subsection{Testing RSD measurements on mocks}

We test the model and our fitting methodology by fitting the average monopole and quadrupole of the correlation function recovered from the 1000 mocks. We use the joint covariance matrix appropriate for a single realisation, including the cross-covariance between the monopole and quadrupole: thus the errors recovered should match those from a single realisation. To perform the fit, we perform a MCMC sampling over models using the publicly available {\sc emcee} python routine \citep{ForemanMackey2013}. For each parameter we quote the best-fit value of the marginalised likelihood, with $1\sigma$ errors defined by the $\Delta \chi^{2} = 1$ regions around this point. Our fiducial fitting choices are as follows: we use $\Delta s=8\mpcoh$ as our fiducial bin width, and keep only those bins with centres $25 \mpcoh \le s \le 160 \mpcoh$. We apply a prior on $\alpha$ based on the results of Paper I, and we apply priors on $\epsilon$ and $\sigma_{8,nl}$ based on results using data from the Planck satellite \citep{PlanckCosmo}. Our fiducial range of scales is chosen based on the facts that including larger scales adds little extra information and the accuracy of the CLPT model starts to decrease below $s = 25\mpcoh$ for the range of halo masses where galaxies in our sample are found \citep{Wang2014}. We will motivate our other choices and demonstrate that our $f\sigma_{8}$ measurements are largely independent of these choices in the following sections .

\begin{table}
\caption{The mean values and one-sigma errors on $f\sigma_{8}$ and $b\sigma_{8}$ from the average of the mocks, recovered from the marginalised probability distribution when different priors are applied and certain parameters are fixed. We expect to recover values $f\sigma_{8}=0.466$ and $1.15 \le b\sigma_{8} \le 1.22$.}
\begin{tabular}{llccc}
\hline
\hline
& \textbf{Average of Mocks:} \\
No. & Case  &  $f\sigma_{8} $ & $b\sigma_{8}$  \\
\hline
1 & Full fit & 					       $0.43^{+0.47}_{-0.32}$ & $1.04^{+0.19}_{-0.18}$ \\[4pt]
2 & Prior on $\alpha$ &                              $0.49^{+0.28}_{-0.29}$ & $1.09^{+0.14}_{-0.19}$ \\[4pt]
3 & Prior on $\sigma_{8,nl}$ &                  $0.45^{+0.19}_{-0.23}$ & $1.19^{+0.12}_{-0.13}$ \\[4pt]
4 & $35 \le s \le 140 \mpcoh$ &                $0.50^{+0.23}_{-0.24}$ & $1.16^{+0.16}_{-0.18}$ \\[4pt]
5 & $\Delta s = 5 \mpcoh$  &                     $0.45^{+0.18}_{-0.22}$ & $1.20^{+0.11}_{-0.13}$ \\[4pt]
6 & $\Delta s = 10 \mpcoh$  &  		       $0.42^{+0.17}_{-0.20}$ & $1.20^{+0.10}_{-0.11}$ \\[4pt]
7 & $\epsilon = 0.00$ &                              $0.50^{+0.13}_{-0.12}$ & $1.18^{+0.10}_{-0.10}$ \\[4pt]
8 & $\alpha = 1.00$, $\epsilon=0.00$ &  $0.50^{+0.13}_{-0.12}$ & $1.18^{+0.08}_{-0.08}$ \\[4pt]
9 & $\alpha = 1.04$, $\epsilon=0.00$ &  $0.52^{+0.13}_{-0.12}$ & $1.24^{+0.08}_{-0.09}$ \\[4pt]
10 & Linear Fit & 				       $0.42^{+0.11}_{-0.11}$ & $1.14^{+0.08}_{-0.08}$ \\[4pt]

\hline
\label{tab:mockrsd}
\end{tabular}
\end{table}

\begin{figure}
\includegraphics[width=84mm]{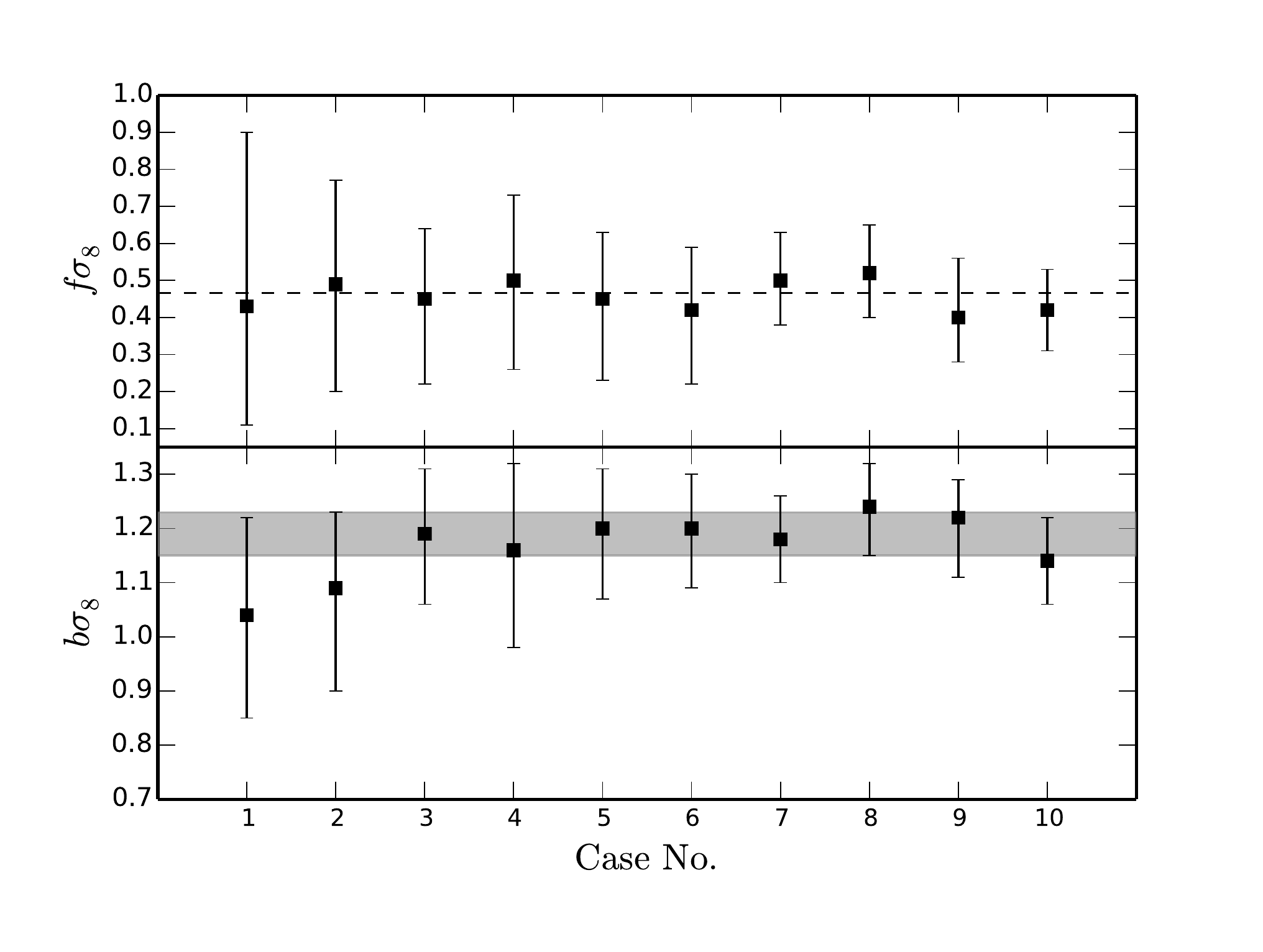}
  \caption{The marginalised $f\sigma_{8}$ and $b\sigma_{8}$ values and one-sigma errors from fitting to the mean of the mocks for the 10 cases listed in Table \ref{tab:mockrsd}. The dashed line indicates the expected growth rate assuming our fiducial $\Lambda$CDM cosmology. The shaded band indicates the expected linear galaxy bias as measured from our HOD fits to the MGS sample, we use a band rather than a line to account for the fact that the calculated value depends slightly on the range of scales used. }
  \label{mockrsdplot}
\end{figure}

The best fit values for all of our fitting cases are collated in Table~\ref{tab:mockrsd}. Fig.~\ref{mockrsdplot} shows the best-fit values for the cases listed in the table along with the $\Lambda$CDM prediction of $f\sigma_{8}$, which closely matches that used in the production of the mock catalogues,  and the expected galaxy bias assuming linear theory \citep{Hamilton1992}. For our fiducial $\Lambda$CDM cosmology, and assuming GR, we have $f(z_{eff})=\Omega_{m}(z_{eff})^{0.55}=0.609$ and $\sigma_{8}(z_{eff}) = 0.766$, and from our HOD fits to the MGS we have $1.5 \le b \le 1.6$ depending on the exact scales used to estimate the linear galaxy bias.

\begin{figure}
\includegraphics[width=84mm]{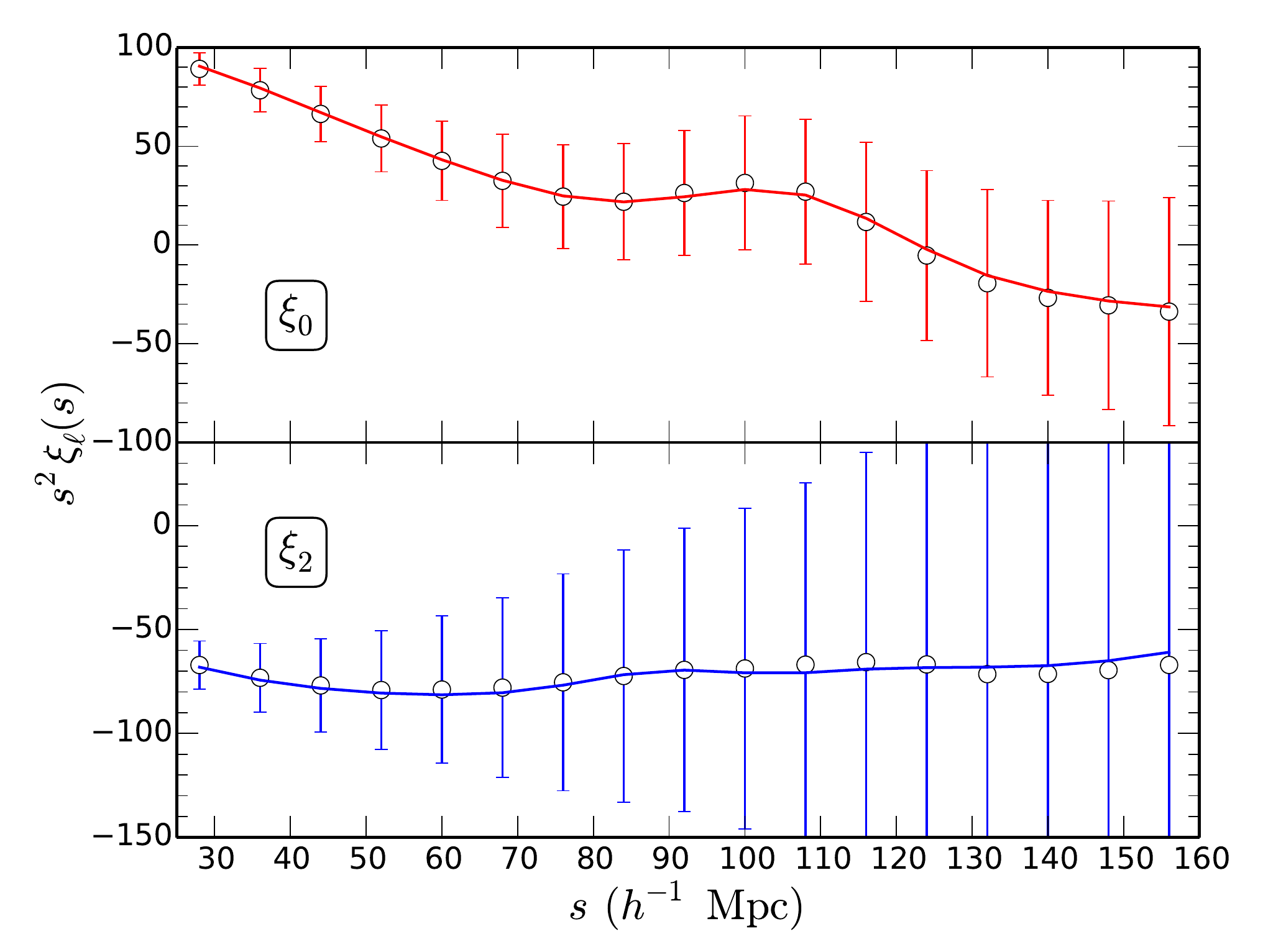}
  \caption{The average monopole and quadrupole of our 1000 mock catalogues (points) shown alongside the best-fit model for our fiducial fitting case (solid) which includes both priors on $\alpha$ and $\sigma_{8,nl}$. The errors are derived from the covariance matrix and are the errors on a single realisation. The CLPT model does a fantastic job of reproducing the measured clustering on all scales of interest.}
  \label{RSDMCmock}
\end{figure}

In Fig.~\ref{RSDMCmock} we plot the best-fitting model monopole and quadrupole for our fiducial fit alongside that measured from the average of mocks. We can see that the CLPT model does remarkably well in modelling the monopole and quadrupole across all the scales we fit against, with only small inaccuracies at the smallest scales and around $s=100\mpcoh$. The inaccuracies are clearly well below the expected level of noise in our measurements.

\subsubsection{Effects of $\alpha$ Prior} \label{sec:alphapriors}

We include a prior on $\alpha$, motivated by the expected improvement in the BAO peak position after reconstruction, in our fiducial $f\sigma_{8}$ measurements, and we test the effect of including this for mock results in this section. Much of the information on $\alpha$ comes from the BAO feature, however in our data, as may be inferred from Fig.~\ref{xi0mock}, the BAO feature in the monopole is very noisy. Reconstruction provides a means for us to recover more of the information within the BAO feature and hence can improve our constraints on $\alpha$, as was done in Paper I. During reconstruction we assume a linear RSD model to convert the galaxy overdensity in redshift space to a Lagrangian displacement for each galaxy. It is common, but not necessary, to also scale the displacements to remove the linear RSD and simplify the BAO constraints by making the amplitude of the signal isotropic when analysed in the true cosmology. The effect of this process on the quadrupole of the correlation function is not well understood and so post-reconstruction measurements cannot currently be used for RSD constraints.

However, as a result of the BAO fits in Paper I, we still have a greater knowledge of $\alpha$ than is apparent in the pre-reconstruction monopole. We encapsulate this using a Gaussian prior on $\alpha$, centred on the recovered post-reconstruction best-fit values from Paper I, and with a variance calculated from the difference between pre- and post-reconstruction fits to the BAO feature (the pre-reconstruction uncertainty is a factor 2.5 times greater than the post-reconstruction result). In other words, we expect the inclusion of the $\alpha$ prior to recover the same uncertainty on $\alpha$ as found in Paper I. Reconstruction also shifts the position of the BAO peak due to the removal of coupling between different k-modes on the scale of the BAO feature. Paper I fits the post-reconstruction (hence no mode-coupling) correlation function with a model that does not include mode-coupling, whereas we fit the pre-reconstruction correlation function with a non-linear model that does include mode-coupling and hence the expected values of $\alpha$ returned by both methods should be the same.

We find that including such a prior has only a small effect on the recovered values and errors for $f\sigma_{8}$ and $b\sigma_{8}$, slightly decreasing the error range for both. The recovered best-fit values only change by a small amount compared to the statistical error on the measurements. This indicates that such a process introduces no bias into our results, which is not surprising, as the $\alpha$ prior comes from the comparison of the data itself before and after reconstruction, and we expect systematic effects entering during the reconstruction process to be very small. The reduction in the error range comes from the improvement in the Alcock-Paczynski measurement when the BAO position is known, and not from double counting as we have carefully only included the extra information recovered post-reconstruction.

\subsubsection{Effects of $\sigma_{8,nl}$ Prior} \label{sec:sigmapriors}

The CLPT model's dependency on $\sigma_{8,nl}$ in the non-linear regime is weak enough that our data provides no constraints on this except through the first order measurements of $b\sigma_{8}$ and $f\sigma_{8}$. The remaining non-linear contribution is largely unconstrained. We therefore consider a Planck+WP+highL prior on $\sigma_{8,nl}$ (Planck Collaboration et al. 2013), which takes the form of a Gaussian with mean $\sigma_{8,nl}(z_{eff})=0.766$ and variance 0.012, so that the second order corrections to the model do not stray into unphysical regions of parameter space, where the model itself is not expected to be accurate. For our baseline fits, we adopt this prior, which we consider not to be introducing any additional information to our measurements, but simply forcing us to only consider physical solutions for the CLPT model.

When we include this prior there is a small change in the recovered mean values of $f\sigma_{8}$ and $b\sigma_{8}$. For the average of the mocks we can see that the value of $f\sigma_{8}$ decreases slightly from 0.49 to 0.45. This shift actually brings the values of $f\sigma_{8}$ closer to that expected based on the cosmology used to generate the mocks and is well within the expected statistical deviation of the measurement. Additionally, adding in the $\sigma_{8,nl}$ prior increases the value of $b\sigma_{8}$ and tightens our constraints, bringing them closer to the expected value. This is because the prior allows us place constraints on the second order contribution to the galaxy bias, which, in the CLPT model, enters as additional small scale clustering proportional to $\langle F'' \rangle^{2}$. When this contribution is completely unconstrained, large values force the linear galaxy bias to be lower than it should be to fit the smallest scales. Due to the strong degeneracy between $b\sigma_{8}$ and $f\sigma_{8}$ it is actually this stronger constraint on $b\sigma_{8}$ that has a knock-on effect of reducing the value of $f\sigma_{8}$ we obtain. 

\subsubsection{Testing bin width and fitting range}

We perform several robustness tests using the $\alpha$ and Planck prior measurement, looking at the effects of changing both the bin width of our measurements and the fitting range. When we change the fitting range to $35 \le s \le 140$ we see a slight increase in $f\sigma_{8}$, and corresponding decrease in $b\sigma_{8}$, though these shifts are well within the statistical uncertainty. The reason for this shift stems from the higher order Lagrangian bias contributions: when we remove the small scale data, our constraints on $\langle F'' \rangle$ become much weaker and it is harder to decouple from $\langle F' \rangle$. We can also see that the errors on $f\sigma_{8}$ and $b\sigma_{8}$  increase when we reduce our fitting range, consistent with the loss of information, particularly at small scales.

The results in Table~\ref{tab:mockrsd} and Figure~\ref{mockrsdplot}  also show that our choice of bin width has negligible effect on the results we obtain. In Cases 5 and 6 we perform fits using our fiducial fitting range and priors but using a correlation function and covariance matrix that has been binned using $\Delta s=5\mpcoh$ and $\Delta s=10\mpcoh$ respectively. We find that the results are fully consistent with each other and our fiducial bin width case, with only small, statistically driven deviations in the mean and $1\sigma$ marginalised values of $f\sigma_{8}$ and $b\sigma_{8}$.

\subsubsection{Effects of Fixing $\alpha$ and $\epsilon$} \label{sec:fixedepsilon}

\begin{figure}
\includegraphics[width=84mm]{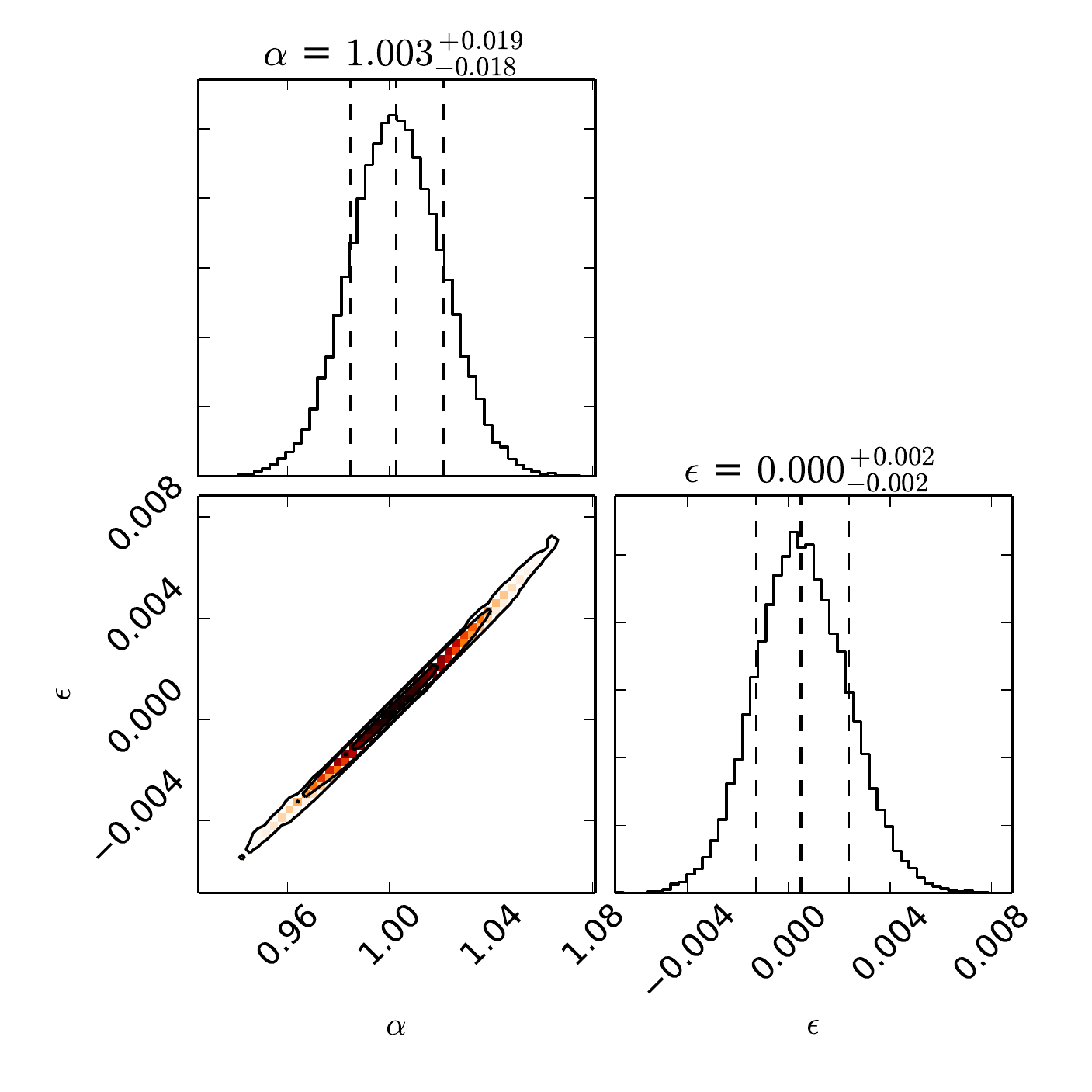}
  \caption{The 2D and 1D marginalised constraints on $\alpha$ and $\epsilon$ at $z=0.15$ based on Planck $\Lambda$CDM cosmological constraints. Ellipses show the 1, 2 and 3$\sigma$ regions, whilst dashed lines show the mean and $1\sigma$ errors of the marginalised distributions.}
  \label{planckAP}
\end{figure}

We also look at models where we do not vary the values of $\alpha$ and $\epsilon$, as in several previous studies \citep{Blake2011a, Beutler12, Samushia2012}. This carries the implicit assumption that our fiducial cosmology is the true cosmology. Figure~\ref{planckAP} shows the expected deviation of these parameters, assuming $\Lambda$CDM, at our effective redshift based on the cosmological results from Planck \citep{PlanckCosmo}\footnote{We used the Planck $\Lambda$CDM base-planck-lowl-lowLike-highL chains which, at the time of writing, are publicly available for download from the Planck Legacy Archive at http://pla.esac.esa.int/pla/aio/planckProducts.html.}, which is the basis for our fiducial cosmology.

We see that $\epsilon$, which is related to the AP parameter F as in Eq. (21), is very well defined at the effective redshift of our sample, with only a 1\% deviation from $\epsilon=0.0$ allowed to within 5$\sigma$. Even relatively large deviations from our fiducial cosmology manifest as only small changes in $\epsilon$ away from zero. As a majority of the information on $\epsilon$ comes from the quadrupole, which is also where we obtain most of the information on $f\sigma_{8}$, we can conclude that the actual AP signal we expect to measure as part of our fitting is also small.

However, from Figure~\ref{planckAP} we can also see that fixing alpha to our fiducial value is not supported by the Planck data, where even large deviations from $\alpha=1.0$ can be found to within 5$\sigma$. It is mainly the monopole of the correlation function that constrains $\alpha$, but the large degeneracies between $\alpha$ and $b\sigma_{8}$, and $b\sigma_{8}$ and $f\sigma_{8}$ means that fixing this value could have a knock-on effect on our $f\sigma_{8}$ constraints. As such we hypothesise that though the expected degeneracy between the AP and RSD signals is small, not allowing $\alpha$ to vary could bias our constraints on $b\sigma_{8}$ and $f\sigma_{8}$.

Finally, it also important to note that Figure \ref{planckAP} is only true when we assume a $\Lambda$CDM cosmology. Allowing for $w_{0} \ne 1.0$, a time-dependent equation of state for dark energy, or other non-standard cosmological models could allow for a much greater variation in $\alpha$ and $\epsilon$ from their fiducial values. As these phenomena are only emergent at late times they would be largely unconstrained by Planck, rendering any apparent Planck priors on $\alpha$ and $\epsilon$ moot.

To test this we perform additional fits to the average of the mocks: first fixing $\epsilon=0.0$ and allowing alpha to vary, then fixing $\epsilon$ and $\alpha$. We fix $\alpha$ to two different values: $\alpha=1.00$, which is what we expect for the mean of the mocks, and $\alpha=1.04$ which is the value recovered from the BAO-only fits to the MGS data in Paper I.

From Table~\ref{tab:mockrsd} and Figure~\ref{mockrsdplot} we can see the recovered values of $f\sigma_{8}$ and $b\sigma_{8}$ when fixing $\epsilon$ do shift slightly, but are still in very good agreement with the expected values for the mocks. This indicates that we are not introducing any bias into our results. The uncertainty on $f\sigma_{8}$ is also reduced substantially, with the lower bound especially reduced by a factor of 2. This is because confining our model to only those regions of parameter space that are in agreement with the Planck-$\Lambda$CDM predictions greatly reduces the degeneracy between $f\sigma_{8}$ and $\epsilon$, improving our constraints.

It should be noted however that this result would also be recovered if we were to take the case where we vary $\alpha$ and $\epsilon$ and simply combined with Planck data at a later stage, as the constraints from Planck are tight enough to effectively fix $\epsilon$. The benefit to allowing $\epsilon$ to vary is that the subsequent $f\sigma_{8}$ results are more general and can be combined with any additional models, not just those that agree with the Planck-$\Lambda$CDM constraints.

When fixing $\alpha$ to different values we do see a small change in the recovered best fit values of $b\sigma_{8}$ and $f\sigma_{8}$, whilst the uncertainties therein remain unchanged. However this is not much beyond that seen when fixing $\epsilon$ to the value expected from the mocks. We will reiterate, however, that fixing $\alpha$ is not supported by the Planck-$\Lambda$CDM predictions and so this should be allowed to vary.

\subsubsection{Using a Linear Model}

Lastly, we investigate the case where we perform a simple linear model fit as per \cite{Hamilton1992}. In Table~\ref{tab:mockrsd} and Figure~\ref{mockrsdplot} we show the results when fitting using a linear model. Here we still keep our reconstruction-motivated prior on $\alpha$, and vary $f\sigma_{8}, b\sigma_{8}, \alpha, \epsilon$ and $IC$. In this case we find that the error budget for both $f\sigma_{8}$ and $b\sigma_{8}$ is significantly reduced in comparison to our fiducial fit, and to a greater extent than when we use our perturbation theory model but fix $\alpha$ and $\epsilon$. A simple linear model neglects the contributions from higher order bias corrections which for our sample are non-negligible and have been shown to affect our estimation of $b\sigma_{8}$ and, by way of the strong degeneracy therein, $f\sigma_{8}$. However, we find that there is no significant bias in the recovered best-fit values themselves when using a linear model and that any differences between the observed RSD signal and the prediction from linear theory are largely hidden by noise.

\section{Results}

\begin{table}
\caption{The mean values and one-sigma errors on $f\sigma_{8}$ and $b\sigma_{8}$ from fitting to the data monopole and quadrupole, when different priors are applied and certain parameter combinations are fixed. From $\Lambda$CDM and GR we expect $f\sigma_{8} = 0.466$ and from our HOD fits to the MGS data we expect $1.15 \le b\sigma_{8} \le 1.22$.}
\begin{tabular}{llccc}
\hline
\hline
& \textbf{Data:} \\
No. & Case  &  $f\sigma_{8} $ & $b\sigma_{8}$ & $\chi^{2}$/dof  \\
\hline
1 & Full fit & 		              	    	     $0.63^{+0.24}_{-0.27}$ & $1.00^{+0.21}_{-0.19}$ & 26.0/26 \\[4pt]
2 & prior on $\alpha$ &  	                        $0.64^{+0.23}_{-0.22}$ & $0.98^{+0.16}_{-0.20}$ & 26.2/26 \\[4pt]
3 & prior on $\sigma_{8,nl}$ &                 $0.53^{+0.19}_{-0.19}$ & $1.17^{+0.14}_{-0.18}$ & 28.6/26 \\[4pt]
4 & $35 \le s \le 140 \mpcoh$ &               $0.56^{+0.25}_{-0.24}$ & $1.08^{+0.14}_{-0.22}$ & 25.8/20 \\[4pt]
5 & $\Delta s = 5 \mpcoh$ &		      $0.52^{+0.19}_{-0.19}$ & $1.16^{+0.13}_{-0.16}$ & 40.1/46 \\[4pt]
6 & $\Delta s = 10 \mpcoh$ & 		      $0.49^{+0.17}_{-0.22}$ & $1.19^{+0.12}_{-0.15}$ & 18.8/20 \\[4pt]
7 & $\epsilon = 0.00$ &                             $0.49^{+0.15}_{-0.14}$ & $1.20^{+0.15}_{-0.15}$ & 31.0/27 \\[4pt]
8 & $\alpha = 1.00$, $\epsilon=0.00$ & $0.44^{+0.16}_{-0.12}$ & $1.12^{+0.09}_{-0.14}$ & 30.3/28 \\[4pt]
9 & $\alpha = 1.04$, $\epsilon=0.00$ & $0.49^{+0.16}_{-0.13}$ & $1.17^{+0.10}_{-0.12}$ & 31.0/28 \\[4pt]
10 & Linear Fit & 				      $0.47^{+0.13}_{-0.13}$ & $1.15^{+0.08}_{-0.08}$ & 31.1/29 \\[4pt]        
\hline
\label{tab:datarsd}
\end{tabular}
\end{table}

\begin{figure}
\includegraphics[width=84mm]{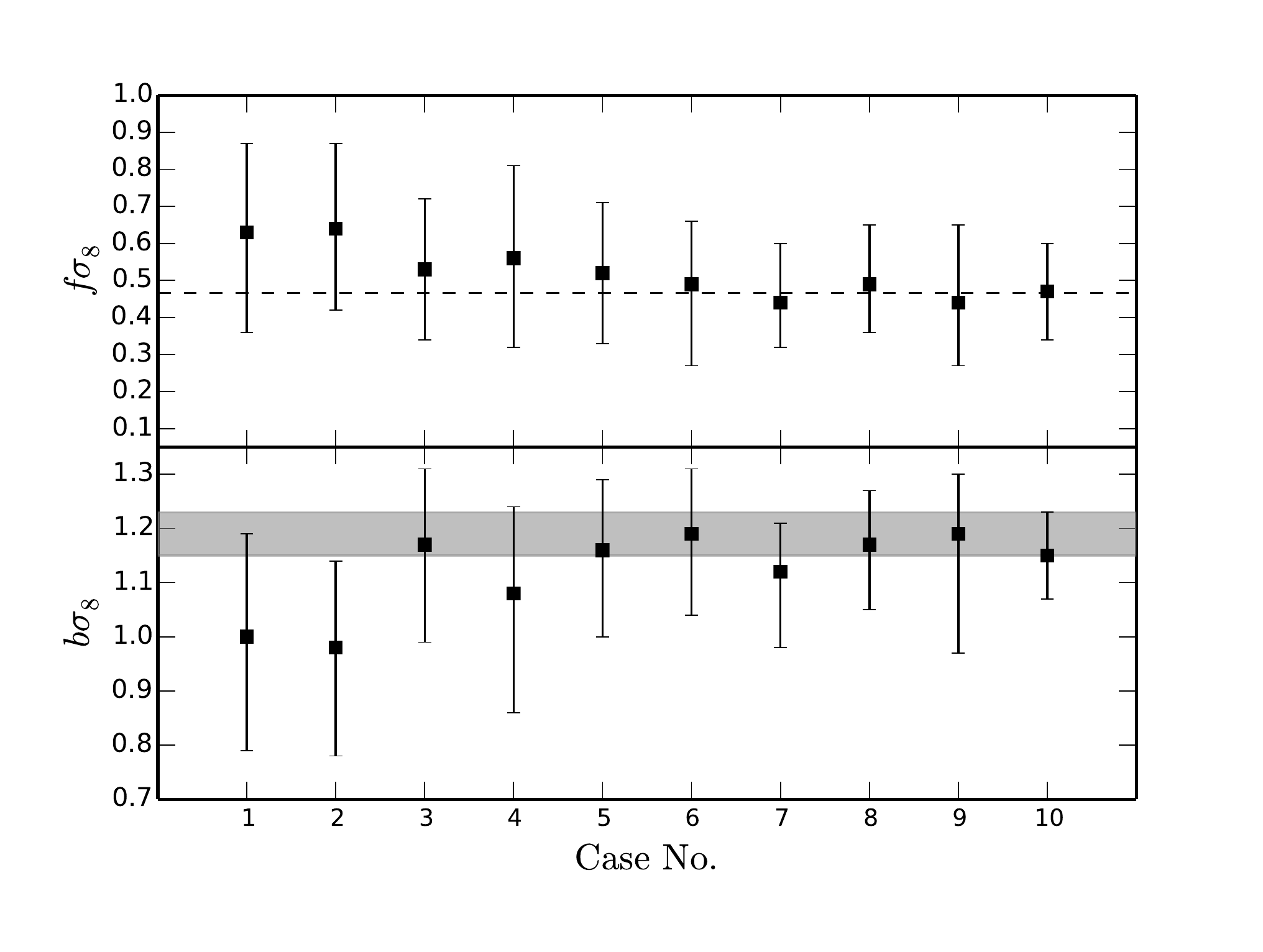}
  \caption{The marginalised $f\sigma_{8}$ and $b\sigma_{8}$ values and one-sigma errors from fitting to the data for the 10 cases listed in Table \ref{tab:datarsd}. As for Fig. \ref{mockrsdplot}, the dashed line indicates the expected growth rate assuming our fiducial $\Lambda$CDM cosmology. The shaded band indicates the expected linear galaxy bias as measured from our HOD fits to the MGS sample, we use a band rather than a line to account for the fact that the calculated value depends slightly on the range of scales used. }
  \label{datarsdplot}
\end{figure}

\begin{figure*}
\includegraphics[width=\textwidth]{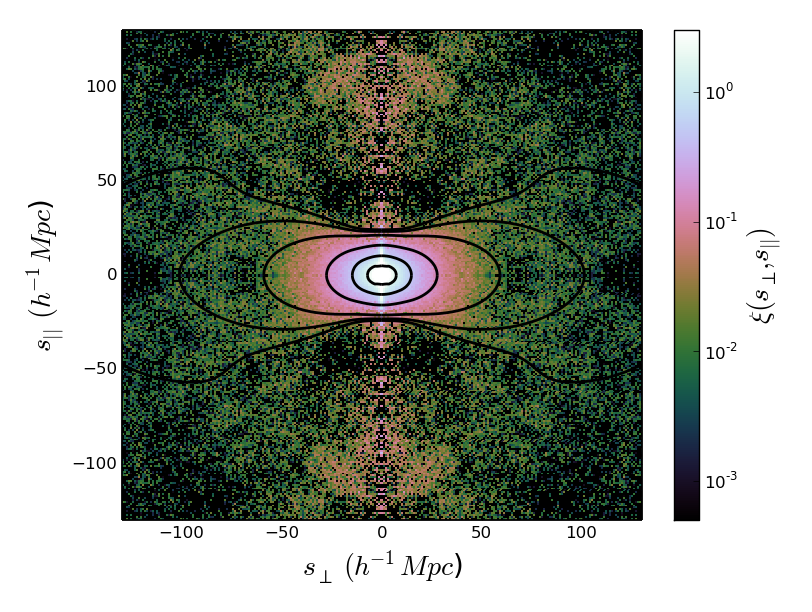}
  \caption{The 2D redshift space correlation function of the MGS along and perpendicular to the line of sight in bins of $\Delta s=1\mpcoh$. The solid black contours show the best-fit CLPT model at $\xi = \{0.001,0.01,0.04,0.3,2.0,15.0\}$ for our fiducial fitting procedure.}
  \label{RSDMC2d}
\end{figure*} 

\begin{figure}
\includegraphics[width=84mm]{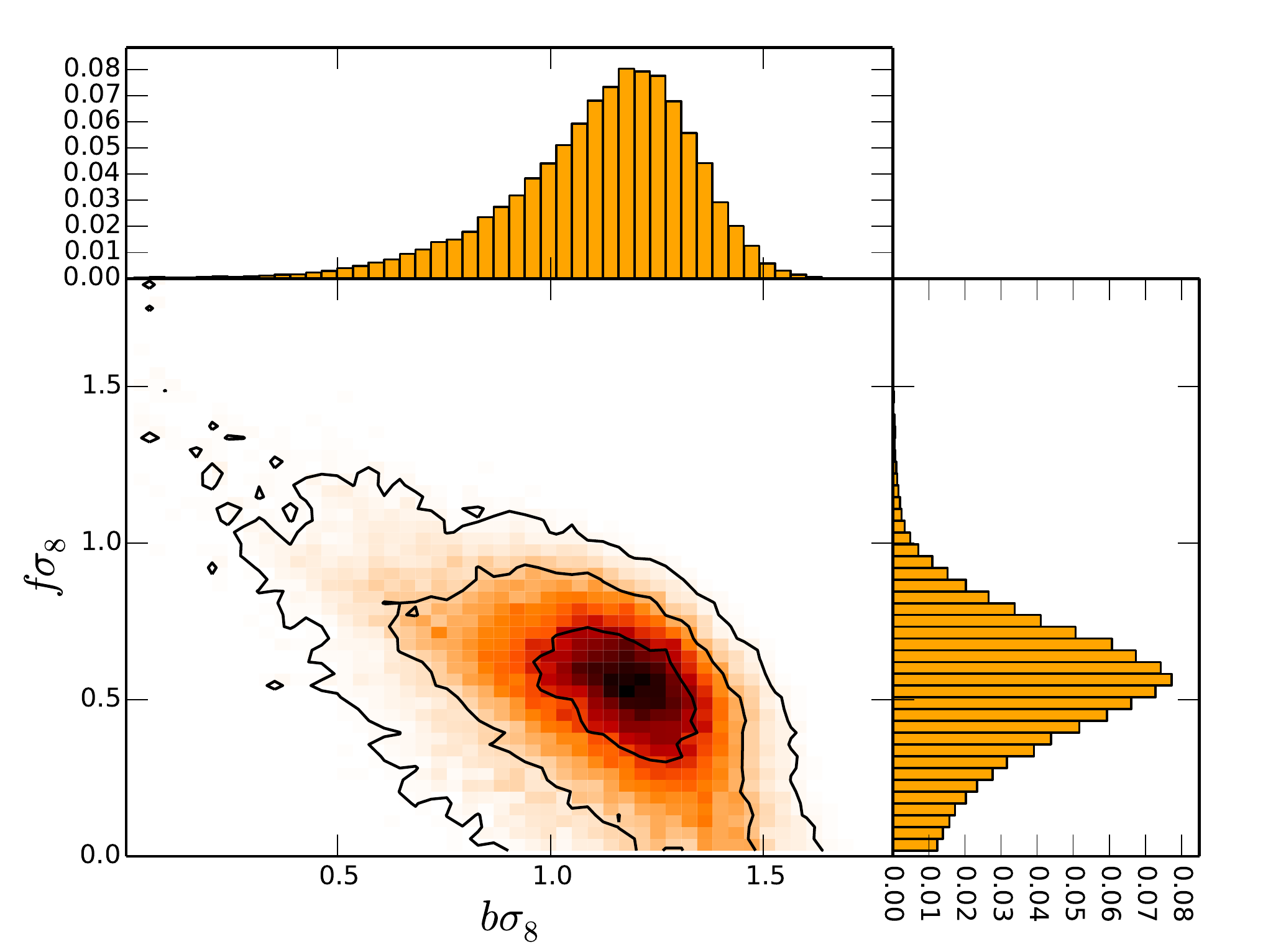}
  \caption{The 1, 2 and 3$\sigma$ $b\sigma_{8}$ and $f\sigma_{8}$ likelihood contours and respective 1D marginalised likelihoods for the MGS galaxy sample using our fits to the monopole and quadrupole in the range $25\mpcoh \le s \le 160\mpcoh$ with bins of width $\Delta s=8\mpcoh$ and priors on $\alpha$ and $\sigma_{8,nl}$.}
  \label{RSDMCcontour} 
\end{figure}

In this section we present our constraints on $f\sigma_{8}$ and $b\sigma_{8}$ from fitting to the MGS data using the method detailed and tested in the previous section. We have shown that our fitting method is independent of our choice of priors, fitting range and bin size, but in the interest of completeness we perform a range of fits equal to those performed on the average of the mocks. For equivalent fits to both data and mocks we use the same covariance matrix, so any differences stem from noise in the data or, of course, differences between our fiducial cosmology and the true cosmology. The marginalised mean values and $1\sigma$ constraints on $f\sigma_{8}$ and $b\sigma_{8}$ for all of our fits are given in Table~\ref{tab:datarsd} with the minimum $\chi^2$ values, and shown in the corresponding Fig.~\ref{datarsdplot}. 

As for the results fitting the average of the mocks, we can see that adding a prior on $\alpha$ introduces no noticeable bias to our best fit $f\sigma_{8}$ and $b\sigma_{8}$ values and only a slight reduction in the errors. When fitting to the data, the best fit $\chi^{2}$ increases slightly from $26.0$ to $26.2$ for 26 degrees of freedom (34 bins and 8 free parameters) when we introduce our prior on $\alpha$. Such an increase is to be expected as the prior forces our best-fit model away from the overall maximum likelihood model, however the difference is very small indicating no strong preference for models outside our prior range. 

When we add in the Planck prior on $\sigma_{8,nl}$ we find a larger difference in the $f\sigma_{8}$ and $b\sigma_{8}$ constraints than for the mocks, though the value of $f\sigma_{8}$ does not shift by more than we would expect based on the statistical errors, and as we do not believe this prior to be adding in any bias to our results from our tests on the mocks, this change is purely statistically driven. Before adding in the $\sigma_{8,n}$ prior the measured values of $b\sigma_{8}$ are lower than we would expect, but  this value increases by $\sim 1\sigma$ when this prior is included. It is this change in the mean recovered value of $b\sigma_{8}$ which causes the slight change in $f\sigma_{8}$. The reason for the underestimation of $b\sigma_{8}$ is as mentioned previously; without this prior helping to constrain $\sigma_{8,nl}$ we overestimate $\langle F'' \rangle$ and hence underestimate $b\sigma_{8}$. For this prior we find $\chi^{2}=28.6$, which is again a slight increase compared to the fits with only the $\alpha$ prior, however for all three cases with different priors the recovered $\chi^{2}$ values for our model are very reasonable.

Our fiducial fitting case including both $\alpha$ and $\sigma_{8,nl}$ priors is shown in Fig.~\ref{RSDMC2d}, where we plot the 2-D redshift space correlation function of our data along with the maximum likelihood model. In Fig.~\ref{RSDMCcontour}, we also plot the recovered $b\sigma_{8}$\hyp{}$f\sigma_{8}$ contour for our fiducial fitting case, alongside the marginalised 1D histograms for these parameters. Here we can see the strong degeneracy between $f\sigma_{8}$ and $b\sigma_{8}$ that drives the small variations we see in our mean values when fitting to both the data and the average of the mocks. 

When we change the fitting range or the bin size, we see similar results as for our fiducial case, and as with the average of the mocks there is no indication that our fitting choices are creating biased results. As for the average of the mocks removing the smallest scales from our fits reduces our recovered $b\sigma_{8}$ value and increases the error, but the mean $f\sigma_{8}$ remains almost unchanged. For all of our tests of bin width and fitting range, we find $\chi^{2}$ values that are in agreement with our fiducial case and which indicate that all of our fits are good. The largest $\chi^{2}$/dof belongs to the case where we modify our fitting range, where we find $\chi^{2} = 25.8$ for 20 degrees of freedom. However, this value is still very good and we would expect a worse $\chi^{2} \approx 17\%$ of the time.

For all our fits to the data it is worth noting that we do seem to fit a slightly lower value for $b\sigma_{8}$ than we would expect based on our HOD fits to the MGS data. Looking back to Fig. \ref{xi0mock} we can see why. The amplitude of the monopole on the scales $25 \le s \le 60$, where most of our information on the linear bias comes from, seems to be slightly lower for the data than for our HOD fit applied to mocks, though when we include scales above and below this range the mock amplitude is well matched. In our fitting we are not including scales below $s=25\mpcoh$, where the mocks and data are in better agreement,  and so it is not surprising the data prefers slightly smaller values of $b\sigma_{8}$.

\begin{figure*}
\centering
\subfloat{\includegraphics[width=0.5\textwidth]{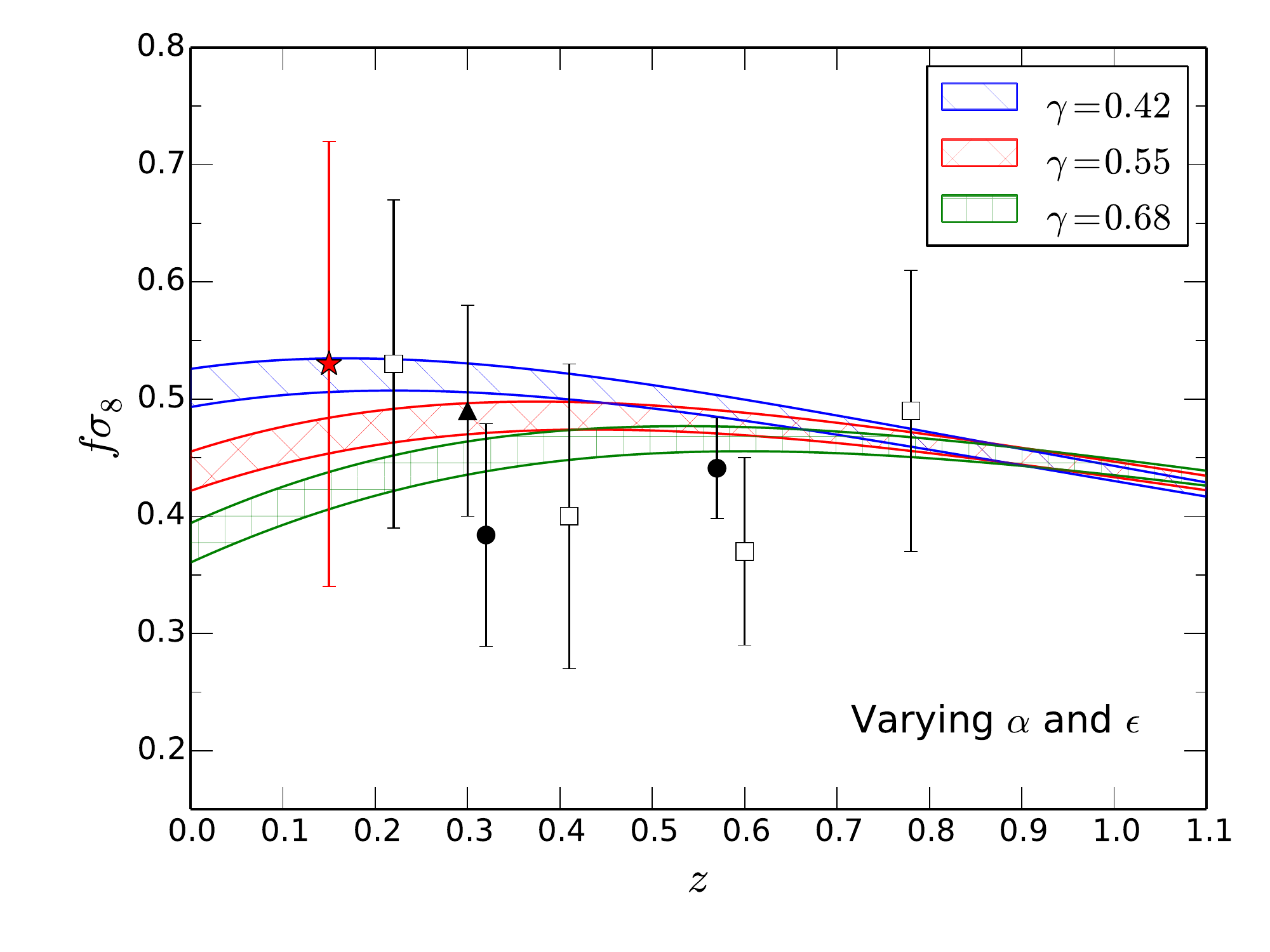}}
\subfloat{\includegraphics[width=0.5\textwidth]{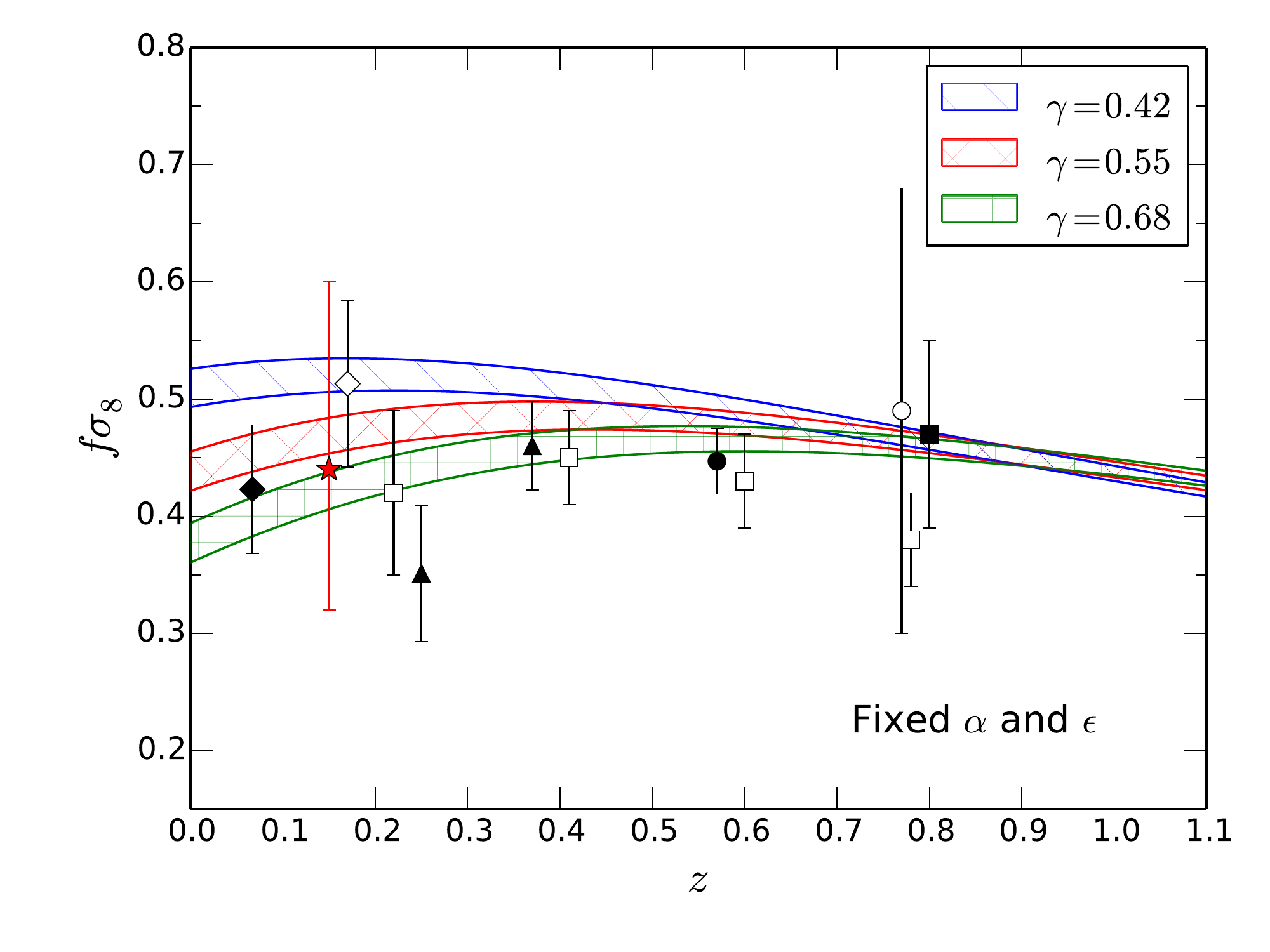}}
\caption{Comparison of measurements of the growth rate using the two-point clustering statistics from a variety of galaxy surveys below $z=0.8$. We split the results into two groups: those that perform a full shape fit, varying $\alpha$ and $\epsilon$; and those that just fit the growth rate for a fixed cosmology, neglecting the degeneracy between $\alpha$, $b\sigma_{8}$ and $f\sigma_{8}$. Our measurement is shown as a filled red star, with other data points representing the 6dFGS (filled diamond; \citealt{Beutler12}), 2dFGRS (empty diamond; \citealt{Percival2004}), SDSS-II LRG (filled triangle; \citealt{Samushia2012} (no AP), \citealt{Oka2014} (AP)), BOSS (filled circle; \citealt{Chuang2013} (z=0.32), \citealt{Samushia2014} (z=0.57)), WiggleZ (open square; \citealt{Blake2011a, Blake2011b}), VVDS (open circle; \citealt{Guzzo2008}) and VIPERS (filled square; \citealt{delaTorre2013}) surveys. We have also included Planck predictions for the growth rate for values of $\gamma = 0.42$, $0.55$ and $0.68$ as hatcheds bands (top, middle and bottom respectively).}
\label{fsigma8plot}
\end{figure*}

The final set of fits we perform, fixing $\alpha$ and $\epsilon$ and using a simpler linear model, corroborate our results when fitting to the average of the mocks. We see that making use of the reasonable assumption that $\epsilon=0.0$ tightens our constraints on $b\sigma_{8}$ and $f\sigma_{8}$ without adding any notable change in the best fit results. The upper and lower bounds on $f\sigma_{8}$ reduce from $0.19$ and $0.19$ to $0.15$ to $0.14$ respectively. Fixing $\alpha$ to different values does change the best fit results slightly too, as was seen in the fits to the mean of the mocks, whilst keeping the errors almost unchanged compared to the fixed $\epsilon$ case. This is not a substantial change, though as we do not have strong Planck constraints on $\alpha$, as we do for $\epsilon$, we conclude that fixing $\alpha$ could lead to biased results.

Overall, the $\chi^{2}$ values we find when fixing $\alpha$ and $\epsilon$ or using a linear model are similar in comparison to using the CLPT model and allowing $\alpha$ and $\epsilon$ to vary. The data is not powerful enough to discriminate between these different models, however from \cite{Wang2014} we do know that we cannot expect that a linear model to fully reproduce the RSD signal on the smallest scales that we fit against, where non-linear effects start to dominate, and that when fitting the RSD signal on these small scales the CLPT model is a more reliable choice. 

\subsection{Comparison of different MGS results}

We have performed several fits to the MGS data assuming different values for $\alpha$ and $\epsilon$. Here we provide an overview of those that we quote, those that should be used for further cosmological studies and those that should not.

By fitting the full-shape of the correlation function monopole and quadrupole, and varying $\alpha$ and $\epsilon$, we find best-fit values of $f\sigma_{8}=0.53^{+0.19}_{-0.19}$ and  $b\sigma_{8}=1.17^{+0.14}_{-0.18}$. These values make no assumption on the underlying, late-time, cosmology and so we recommend the usage of these for future cosmological constraints. In the following section we will use these results to constrain the growth index, $\gamma$, and compare this to the prediction from General Relativity. As the 1-D $f\sigma_{8}$ and 3-D $f\sigma_{8}, \alpha$ and $\epsilon$ likelihoods cannot be well approximated by a Gaussian we use the likelihoods themselves to achieve this, rather than just the quoted numbers. For future analyses making use of our results the prepared MCMC samples for this fit will be made publicly available upon acceptance.

If we assume a $\Lambda$CDM cosmology, we are able to improve our constraints by fixing $\epsilon=0.0$ yet still allowing $\alpha$ to vary. Here we find $f\sigma_{8}=0.49^{+0.15}_{-0.14}$ and  $b\sigma_{8}=1.20^{+0.15}_{-0.15}$. This is well motivated by the Planck data, where we find that, unless we have a late time dark energy model quite different from those commonly assumed, we would expect to detect no deviation from $\epsilon=0.0$. As such this measurement is presented as our quoted, fiducial results and should be used for comparison with other $f\sigma_{8}$ results under the $\Lambda$CDM framework. However, this result should not be combined with Planck data as that would result in effectively double counting the Planck constraints. Rather, from Figure~\ref{planckAP}, we can see that combining our publicly available chains with Planck data will effectively fix $\epsilon$ and recover the fiducial results. From the same figure though we would we not recommend the usage of our results where $\alpha$ is not allowed to vary. In fact, as $\alpha$ dilates the whole correlation function, not just the BAO peak, and captures the late-time cosmological dependence of the shape of the correlation even on small scales, we would recommend that $\alpha$ be allowed to vary for any measurements of the growth of structure.

\section{Cosmological Interpretation and Comparison to Previous Studies}

In this section we compare our measurements of $f\sigma_{8}$ to those from a range of different galaxy surveys and perform a simple consistency test against the prediction of the growth rate from General Relativity (GR) using the commonly used $\gamma$ parameterisation of the growth rate, where $f(z)$ is approximated as
\begin{equation}
f(z) = \Omega_{m}(z)^{\gamma}.
\end{equation}
For GR we have $\gamma \approx 0.55$ \citep{Linder2007}.

Measurements of $f\sigma_{8}$ have been made up to $z=0.8$ using data from the 2-degree Field Galaxy Redshift (2dFGRS; \citealt{Percival2004}), 6-degree Field Galaxy (6dFGS; \citealt{Beutler12}), SDSS-II Luminous Red Galaxy \citep{Samushia2012,Oka2014}, BOSS \citep{Chuang2013, Samushia2014,Sanchez2014,Beutler13}, VVDS \citep{Guzzo2008} and WiggleZ \citep{Blake2011a, Blake2011b} surveys among others. Although these measurements were all made using different models of varying complexity and different fitting methods to either the correlation function or power spectrum, they can be roughly grouped into two distinct categories: those that were made assuming a fixed fiducial cosmological model and those that fit the full shape of the galaxy clustering statistics. The latter simultaneously measures both the RSD and BAO signals and as such includes the degeneracy between $f\sigma_{8}$, $b\sigma_{8}$ and $\alpha$ highlighted in Section \ref{sec:fixedepsilon}

We plot these two sets of measurements separately in Fig. \ref{fsigma8plot}. The $z=0.57$ BOSS and four WiggleZ measurements were calculated with and without the inclusion of the AP effect and we can see that they too find a large difference in the constraints when incorporating this degeneracy into their measurements. Alongside these measurements we also plot the Planck-$\Lambda$CDM predictions for $f\sigma_{8}$ assuming different values for the $\gamma$ parameter. We can see that the majority of the measurements, including our MGS measurements, are in good agreement with the GR prediction.

As a more quantitative consistency test of GR we use the likelihood recovered from our full-fit MCMC analysis to put constraints on $\gamma$ itself. We use our data in combination with the publicly available Planck likelihood chains, subsampling these to enforce a prior on $\Omega_{m}$. We importance-sample the Planck chain by randomly choosing a value $0 \le \gamma \le 1.5$ for each point in the chain and evaluating the likelihood for that parameter combination. One caveat, however, is that we have to correct the value of $\sigma_{8}$ to account for the fact that this also depends on $\gamma$. For each point in the Planck chains we have $\Omega_{m,0}$ and $\sigma_{8,0}$, where the later is derived from the CMB power spectrum amplitude assuming GR. The correct value of $f\sigma_{8}$ is then evaluated by scaling back $\sigma_{8}$ to a suitably high redshift (for simplicity we use the redshift of recombination, $z*$) and then scaling both $\sigma_{8}$ and $\Omega_{m}$ to our effective redshift using the correct value of $\gamma$. i.e., for scale factor $a = 1/(1+z)$, 
\begin{equation}
f(a)\sigma_{8}(a) = \Omega_{m}(a)^{\gamma}\sigma_{8,0}\frac{D_{gr}(a*)}{D_{gr,0}}\frac{D_{\gamma}(a)}{D_{\gamma}(a*)}
\end{equation}
where,
\begin{equation}
\Omega_{m}(a) = \frac{\Omega_{m,0}}{a^{3}E(a)^{2}}
\end{equation}
\begin{equation}
D_{gr}(a) = \frac{H(a)}{H_{0}}\int_{0}^{a}\frac{da'}{a'^{3}H(a')^{3}}
\end{equation}
\begin{equation}
\frac{D_{\gamma}(a)}{D_{\gamma}(a*)} = \text{exp}\left[ \int_{a*}^{a} \Omega_{m}(a')^{\gamma} d\text{ln}a' \right]
\end{equation}
and
\begin{equation}
H(a) = H_{0}E(a) = H_{0}\sqrt{\frac{\Omega_{m,0}}{a^{3}}+\frac{(1-\Omega_{m,0}-\Omega_{\Lambda,0})}{a^{2}}+\Omega_{\Lambda,0}}
\end{equation}

\begin{figure}
\includegraphics[width=84mm]{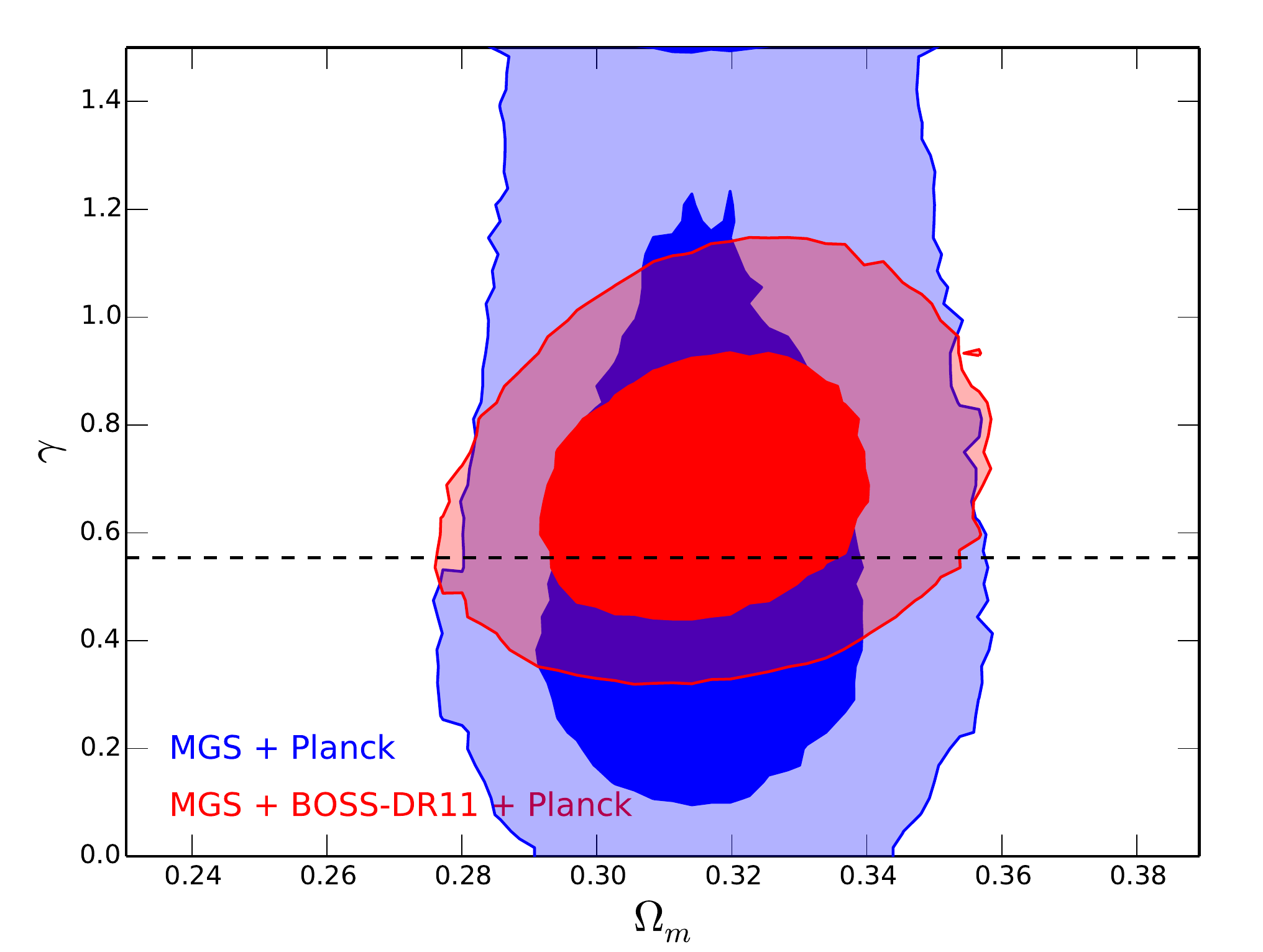}
  \caption{Constraints on $\gamma$ and $\Omega_{m}$ from the combination of our marginalised $f\sigma_{8}$ and Planck likelihoods. Contours correspond to the $1\sigma$ and $2\sigma$ confidence intervals of the recovered posterior distribution. We additionally look at the case where we include the BOSS-DR11 CMASS measurement of the growth rate \citep{Samushia2014}. In both cases we find good agreement with the prediction from GR (dotted line).}
  \label{gamma1d} 
\end{figure}

Even though our fiducial $f\sigma_{8}$ measurements use a prior to better constraint $\sigma_{8,nl}$ and stop the non-linear aspects of the CLPT model from straying into non-physical regions of our cosmological parameter space, all of the information on $f\sigma_{8}$, $\alpha$ and $\epsilon$ comes solely from the amplitude and BAO features of the monopole and quadrupole. As such we are able to combine our results with Planck data for this consistency test without the risk of double counting the Planck measurements.

Our subsequent constraints on $\gamma$ and $\Omega_{m}$ are shown in Fig.~\ref{gamma1d}.  Here we also show the joint constraints when including the measurements of $f\sigma_{8}$ from the BOSS-DR11 CMASS sample \citep{Samushia2014}. For our simple consistency check we only include the CMASS measurement as the method used to make this measurement is very similar to that used in this work. On top of this, the BOSS-DR11 LOWZ and WiggleZ measurements do overlap partially in terms of area and redshift distribution with both our measurement and the CMASS measurement, so to properly include these would require an accurate computation of the cross correlation between these measurements which is beyond the scope of this work. When combining the MGS result with our Planck prior we recover $\gamma=0.58_{-0.30}^{+0.50}$, consistent with GR. With the addition of the CMASS measurement we recover $\gamma = 0.67_{-0.15}^{+0.18}$, which is also consistent with GR to within $1\sigma$. However it should be noted that in both cases we do find a slight preference for higher values of $\gamma$ than would be expected from GR.

\begin{figure}
\includegraphics[width=84mm]{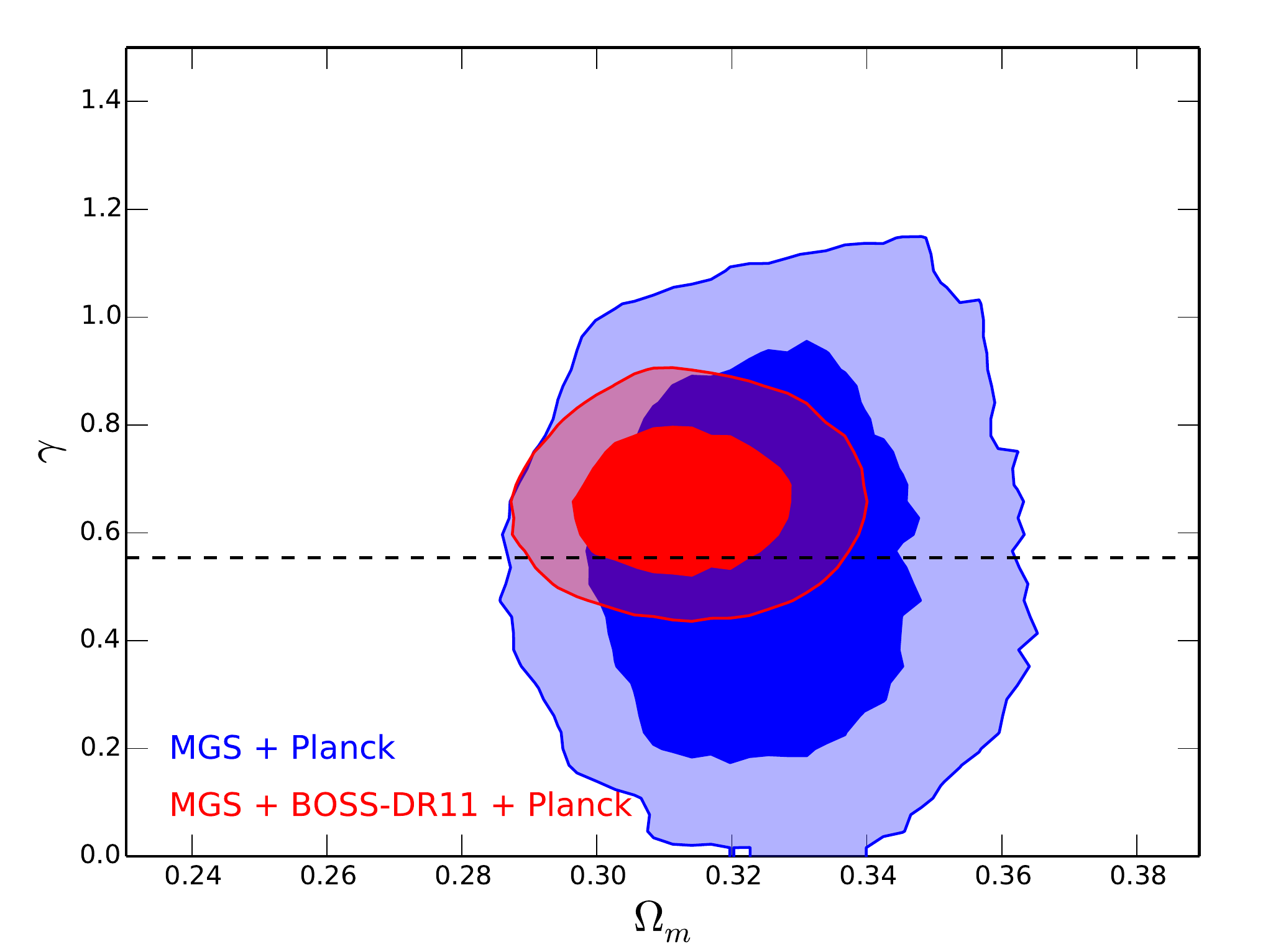}
  \caption{Constraints on $\gamma$ and $\Omega_{m}$ from the combination of our 3-dimensional, marginalised $f\sigma_{8}$, $\alpha$ and $\epsilon$ likelihood with the Planck likelihood. Contours correspond to the $1\sigma$ and $2\sigma$ confidence intervals of the recovered posterior distribution. In both cases we find good agreement with the prediction from GR (dotted line) and a reduction in the uncertainty on $\gamma$, compared to Fig.~\ref{gamma1d}, when we include the anisotropic BAO information from the CMASS and MGS measurements.}
  \label{gamma3d} 
\end{figure}

We take this one step further and include BAO information from our measurement and from the BOSS-DR11 CMASS results as the inclusion of anisotropic distance information helps to better constrain $\Omega_{m}$ and hence can reduce the uncertainty on our $\gamma$ constraints. We use the 3D $f\sigma_{8}, \alpha$ and $\epsilon$ likelihood from our fiducial fits as well as the equivalent constraints from the CMASS sample.  The results of this are shown in Fig. \ref{gamma3d} where we find $\gamma= 0.64 \pm 0.09$ with, and $\gamma = 0.54_{-0.24}^{+0.25}$ without, the inclusion of the CMASS measurement. Both of these measurements are consistent with GR to within $1\sigma$. The addition of our MGS $f\sigma_{8}, \alpha$ and $\epsilon$ measurements improves the constraints on $\gamma$ by $\sim10\%$ compared to the constraints we get on $\gamma$ using the CMASS measurement alone.

\begin{figure}
\includegraphics[width=84mm]{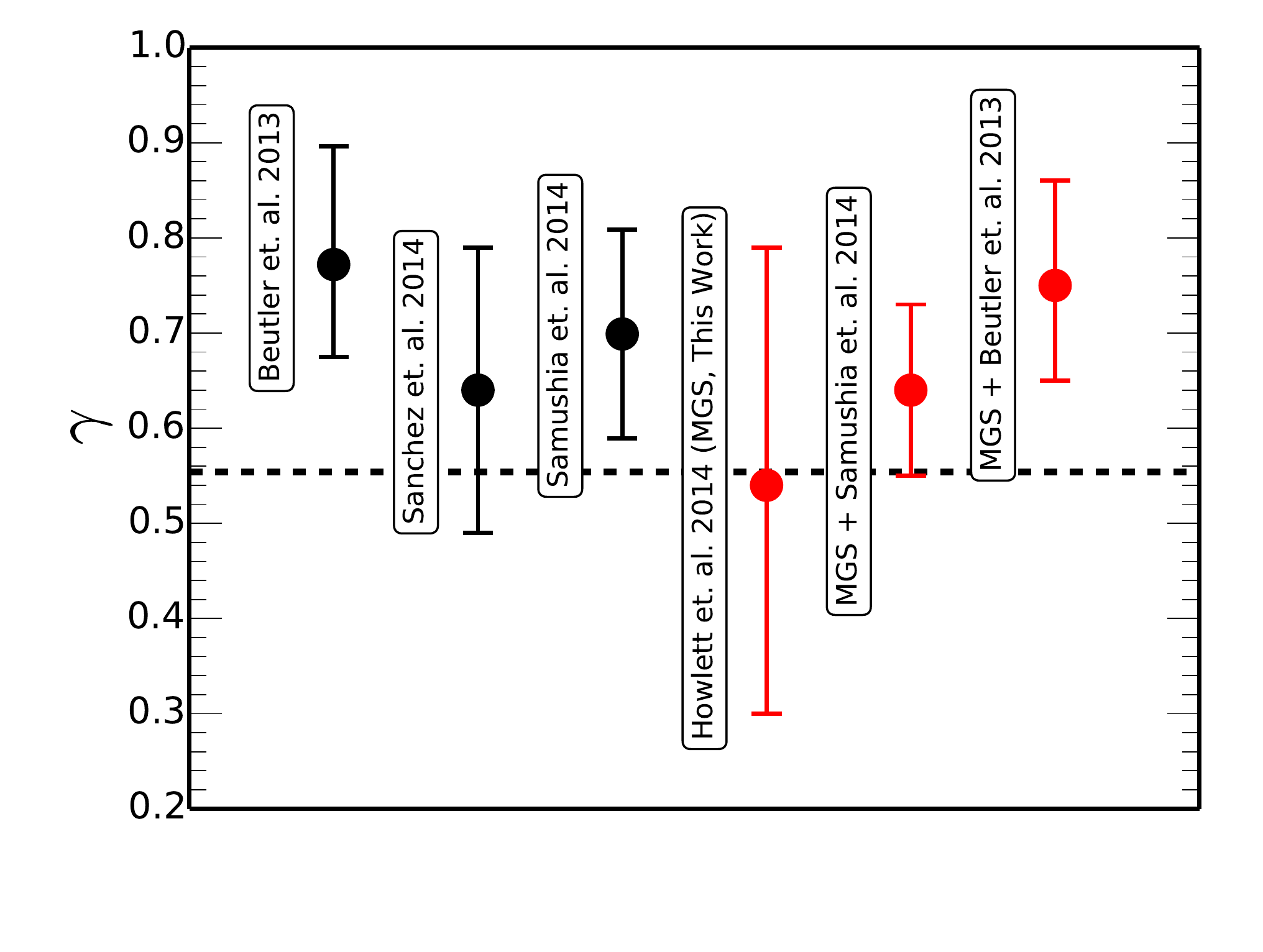}
  \caption{A comparison of $\gamma$ constraints from several independent measurements of the growth rate using combinations of BOSS CMASS (and in the case of \protect\citealt{Sanchez2014}, BOSS LOWZ) and Planck data. For consistency we plot our MGS+Planck only measurement alongside. We can see good agreement between all independent probes and a somewhat consistent favour for higher values of $\gamma$ than would be predicted by GR (dashed line).}
   \label{gamma_compare} 
\end{figure}

The growth index has also been measured by \cite{Beutler13}, \cite{Sanchez2014} and \cite{Samushia2014} from the combination of BOSS CMASS and Planck data. Additionally \cite{Sanchez2014} use BOSS LOWZ data to produce their constraints. In Fig. \ref{gamma_compare} we plot our MGS+Planck constraint on $\gamma$ alongside these other measurements. We see good consistency between all measurements, even though the methods used to measure the growth rate and anisotropic BAO information are very different. In all cases we also see a slight preference for higher values of $\gamma$, which corresponds to models where gravitational interactions are weaker.

There exists significant tension ($\sim 2.3\sigma$) between the \cite{Beutler13} BOSS CMASS measurement of the growth index and  the prediction from GR. An interesting question to ask is whether the addition of our measurements at low redshift helps to alleviate this tension and how this combination of measurements compares to the result presented previously when we combine the MGS and \cite{Samushia2014} BOSS CMASS measurements.  The results from these two combinations are also presented in Fig. \ref{gamma_compare}, where we find that our measurement brings both combinations towards better agreement with the GR prediction, however there is still a $2\sigma$ tension between this prediction and the value of $\gamma$ recovered when combining our measurements with the \cite{Beutler13} CMASS results.

\section{Conclusions}

In this paper we have presented measurements of the growth rate of structure at an effective redshift of $z=0.15$ from fits to the monopole and quadrupole of the correlation function of the SDSS Data Release 7 Main Galaxy Sample (MGS). We have also described the creation of a large ensemble of 1000 simulated galaxy catalogues which enabled both this measurement and the isotropic BAO measurements made in Paper I, where the sample itself is detailed. Our main results can be summarised as follows:
\begin{itemize}
\item{We have used a newly developed code {\sc picola} to generate 500 unique dark matter realisations. We use the Friends-of-Friends algorithm to create halos and populate these halos using a HOD model fitted to the power spectrum of the MGS. We find that the resultant 1000 galaxy catalogues are highly accurate, reproducing the observed clustering down to scales less than $10\mpcoh$. Full details of our code {\sc picola} can be found in Howlett et. al. (in prep.)}
\item{Using these mock catalogues we construct covariance matrices for our two-point clustering measurements and test some of the assumptions made in the BAO fits presented in Paper I. We find: negligible cross-correlation between mock galaxy catalogues generated from the same dark matter field; that the method used to generate our random data points introduces no significant systematic effects; and that we can assume our errors on the power spectrum and correlation function are drawn from an underlying multivariate Gaussian distribution.}
\item{We use the CLPT model \citep{Wang2014} to fit the monopole and quadrupole of the correlation function. We use our mock catalogues to test the model for systematic effects and find excellent agreement between the model and the average monopole and quadrupole of the correlation function. We also perform a series of robustness tests of our method, looking at our choice of priors, fitting range and binsize. In all cases we see no evidence that our results are biased in any way, with all methods recovering the expected value of $f\sigma_{8}$ for our mock catalogues.}
\item{Fitting to the MGS data we measure $f\sigma_{8} = 0.53_{-0.19}^{+0.19}$ when fitting to the full shape of the correlation function and $f\sigma_{8} = 0.49_{-0.14}^{+0.15}$ when assuming no AP effect and fixing $\epsilon=0$. This assumption is validated by the fact that we expect to detect $\epsilon=0$ for any commonly assumed model of the expansion history based on the Planck-$\Lambda$CDM results. However, we have also shown that even at the low effective redshift of our measurement, and assuming $\Lambda$CDM, $\alpha$ can be expected to vary substantially from that expected for our fiducial cosmology. As such, fixing this to a specific value is not recomended for measurements of the growth of structure.}
\item{Using our fiducial results to fit the growth index, $\gamma$, we find $\gamma=0.58_{-0.30}^{+0.50}$ when including Planck data and $\gamma = 0.67_{-0.15}^{+0.18}$ when also including BOSS-DR11 CMASS measurements of the growth rate. When we include the additional anisotropic BAO from the full fits to the shape of the correlation function our constraints tighten to $\gamma= 0.54_{-0.24}^{+0.25}$ and $\gamma = 0.64 \pm 0.09$ respectively, the latter of which is a $\approx 10\%$ improvements on the constraints from the CMASS and Planck measurements alone. All of our results are fully consistent with the predictions of General Relativity, $\gamma \approx 0.55$, and the constraints from other measurements at different redshifts. The MCMC chains used for this analysis will be made publicly available upon acceptance.}
\end{itemize}

\section*{Acknowledgements}

CH is grateful for funding from the United Kingdom Science \& Technology Facilities Council (UK STFC). AJR is thankful for support from University of Portsmouth Research Infrastructure Funding. LS is grateful to the European Research Council for funding. WJP acknowledges support from the UK STFC through the consolidated grant ST/K0090X/1, and from the European Research Council through grants MDEPUGS, and {\it Darksurvey}. 

Mock catalog generation, correlation function and power spectrum calculations, and fitting made use of the facilities and staff of the UK Sciama High Performance Computing cluster supported by the ICG, SEPNet and the University of Portsmouth.

Funding for the creation and distribution of the SDSS Archive
has been provided by the Alfred P. Sloan Foundation, the Participating
Institutions, the National Aeronautics and Space Administration,
the National Science Foundation, the U.S. Department of
Energy, the Japanese Monbukagakusho, and the Max Planck Society.
The SDSS Web site is http://www.sdss.org/.

The SDSS I and II is managed by the Astrophysical Research Consortium
(ARC) for the Participating Institutions. The Participating
Institutions are the University of Chicago, Fermilab, the Institute
for Advanced Study, the Japan ParticipationGroup, Johns Hopkins
University, the Korean Scientist Group, Los Alamos National Laboratory,
the Max Planck Institute for Astronomy (MPIA), the Max
Planck Institute for Astrophysics (MPA), New Mexico State University,
the University of Pittsburgh, the University of Portsmouth,
Princeton University, the United States Naval Observatory, and the
University of Washington.

This research has made use of NASA's Astrophysics Data System Bibliographic Services.

\label{lastpage}

\end{document}